\documentclass[12pt]{article}
\pdfoutput=1
\usepackage{amsfonts}
\usepackage{mathrsfs}
\usepackage{amsmath}
\usepackage{xcolor}
\usepackage{bm}
\usepackage{titlesec}
\usepackage{multirow}
\usepackage{bigstrut}
\usepackage{ulem}
\usepackage[labelfont=bf]{caption}
\usepackage{color}
\definecolor{red}{rgb}{0.75,0,0}
\definecolor{blue}{rgb}{0,0,0.75}
\definecolor{green}{rgb}{0,0.5,0}

\newcommand{\red}[1]{{\color{black} #1}} 
\newcommand{\mcm}[1]{{\color{black} #1}}
\newcommand{\MCM}[1]{{\color{black} #1}}

\newcommand{\yzh}[1]{{\color{black} #1}}
\newcommand{\new}[1]{{\color{black} #1}}

\newcommand{\beginsupplement}{%
        \setcounter{table}{0}
        \renewcommand{\thetable}{S\arabic{table}}%
        \setcounter{figure}{0}
        \renewcommand{\thefigure}{S\arabic{figure}}%
     }

\usepackage{etoolbox}

\DeclareMathOperator{\tr}{tr}

\usepackage{scicite}
\usepackage{times}
\usepackage{graphicx}

\newenvironment{sciabstract}{%
\begin{quote} \bf}
{\end{quote}}

\title{Title: Dynamics of active liquid interfaces}

\author{\textbf{One Sentence Summary:} Active interfaces exhibit giant fluctuations and\\
propagating waves that power wetting transitions and droplet shape-shifting\\
\\
\textbf{Authors:} Raymond Adkins$^{1\dagger}$, Itamar Kolvin$^{1\dagger\ast}$, Zhihong You$^{1\dagger}$, Sven Witthaus,$^{1}$\\
M. Cristina Marchetti$^{1,2\ast}$ and Zvonimir Dogic$^{1,2\ast}$\\
\\
\normalsize{\textbf{Affiliations:}  $^{1}$Department of Physics, University of California at Santa Barbara,}\\
\normalsize{Santa Barbara, CA 93106, USA,}\\
\normalsize{$^{2}$Graduate program in Biomolecular Science and Engineering,}\\
\normalsize{University of California at Santa Barbara, Santa Barbara, CA 93106, USA.}\\
\\
\normalsize{$^\dagger$These authors contributed equally }\\
\normalsize{$^{\ast}$Corresponding authors: zdogic@physics.ucsb.edu (Z.D.),}\\
\normalsize{cmarchetti@ucsb.edu (M.C.M.), itamar@ucsb.edu (I.K.)}}

\date{}

\begin{document} 

\baselineskip24pt

\maketitle 
\begin{sciabstract}
\textbf{Abstract:} Controlling interfaces of phase separating fluid mixtures is key to creating diverse functional soft materials. Traditionally, this is accomplished with surface-modifying chemical agents. Using experiment and theory, we study how mechanical activity shapes soft interfaces that separate an active and a passive fluid. Chaotic flows in the active fluid give rise to giant interfacial fluctuations and non-inertial propagating active waves. At high activities, stresses disrupt interface continuity and drive droplet \red{generation, producing} an emulsion-like active state comprised of finite-sized droplets. When in contact with a solid boundary, active interfaces exhibit non-equilibrium wetting transitions, wherein the fluid climbs the wall against gravity. These results demonstrate the promise of mechanically driven interfaces for creating a new class of soft active matter.
\end{sciabstract}

\textbf{Main Text:} Liquid-liquid phase separation (LLPS) is a ubiquitous phase transition, with examples abounding throughout material science, biology and everyday life~\cite{Bray2002,Brangwynne2009}. Immiscible liquid phases are separated by sharp, but deformable, interfaces that strongly couple to flows and the input of mechanical energy. For example, gentle shaking of an oil-water mixture induces gravity-capillary interfacial waves, while more vigorous perturbations break up the entire interface, reinitializing the phase separation~\cite{Hinze1955,Tong1989,Berthier2001,Perlekar2014}. Active matter provides an alternative method of continuously stirring a fluid~\cite{Simha2002,Marchetti2013}. In such systems, mechanical energy, inputted locally through the motion of microscopic constituents, cascades upward to generate large-scale turbulent-like dynamics~\cite{Sanchez2012,Wensink2012,Zhou2014}. We study how active stresses and associated flows perturb soft interfaces and LLPS. Using experiment and theory, we identify universal features of active-LLPS, including giant interfacial fluctuations, traveling interfacial waves, activity-arrested phase separation  and activity-induced wetting transitions. These results demonstrate how active matter drives liquid interfaces to configurations that are not accessible in equilibrium. In turn, active interfaces are elastic probes that provide insight into the forces driving active fluids, for example by allowing for the measurement of the active stresses.

The active liquid interfaces studied here belong to a wider class of activity-driven boundaries, that includes lipid bilayers, colloidal chiral fluids and interfaces between motile and immotile bacteria in a swarm \cite{Soni2019,Soni2019a,Patteson2018,Takatori2020,Vutukuri2020}. From a biology perspective, LLPS has emerged as a ubiquitous organizational principle  \cite{Brangwynne2009,Caragine2019}. How cytoskeletal active stresses couple to self-organization of membraneless organelles remains an open question. Studies of simplified systems can shed light on these phenomena. Relatedly, active wetting plays a potential role in the development and shaping of tissues~\cite{Perez-Gonzalez2019}. 

To explore how activity modifies soft interfaces, we combined poly(ethylene-glycol) (PEG) and polysaccharide dextran with stabilized microtubule filaments and clusters of kinesin molecular motors. Above a critical polymer concentration, the passive PEG-dextran mixture phase separated~\cite{Liu2012}. Microtubules and kinesin clusters exclusively partitioned into the dextran phase, where depletion forces promoted microtubule bundling (Fig. 1A-C). Streptavidin-bound kinesin clusters (KSA) stepped along adjacent microtubules within a bundle, driving interfilament sliding. The kinesin-powered bundle extensions continuously reconfigured the filamentous network, generating large-scale turbulent-like flows, similar to those previously studied~\cite{Sanchez2012} (Fig. 1D). The PEG-dextran interfaces were susceptible to large deformations by active stresses generated within the dextran phase, due to their ultra-low interfacial tension ($<1$~$\mu\mathrm{N/m}$)~\cite{Liu2012}. 

We first visualized the phase separation dynamics of active-LLPS in $\sim$30 $\mathrm{\mu m}$ thick horizontal microscopy chambers. In such samples, PEG-dextran interfaces had a nearly-flat vertical profile (Fig. S1). The quasi-2D nature of the system was supported by a nearly constant area fraction of the PEG-rich domains (Fig. S2). In a passive system with microtubules but no kinesin motors, the droplets coalesced slowly (Fig. 1E, Movie S1). The addition of motors altered the coarsening kinetics. At \red{intermediate} KSA concentrations, active flows powered droplet motility, which increased the probability of droplets encountering each other and coalescing, thus speeding up coarsening dynamics (Fig. 1E, Movie S2). Higher KSA concentrations \red{accelerated buildup of interfacial fluctuations leading to} an entirely different dynamical state where \red{droplets incessantly fused and fissioned with each other} (Fig\red{s}. 1E, 1G, Movie S3).

To \red{quantify} the influence of activity on the PEG-dextran phase separation, we measured the equal-time two-point correlation function $C(\mathbf{\Delta r},t) = \langle I(\mathbf{r} +\mathbf{\Delta r},t)I(\mathbf{r},t)\rangle_\mathbf{r}$, where $I=1$ in the dextran phase and $-1$ otherwise (Fig. S3). Spatial correlations decayed over a length scale $\xi$, defined by $C(\xi)=0.5$, which is comparable to the average droplet size (Fig. S4). For passive samples, $\xi$ increased slowly in time (Fig. 1F).  Enhanced coarsening at \red{intermediate} KSA concentration was reflected by a much faster initial growth of $\xi$ than the passive case. At high motor concentration, $\xi$ peaked at $\sim$\red{1} hour and subsequently decayed to a finite plateau, $\xi_{steady}$. In parallel, average interface curvature $\kappa$ monotonically grew, reaching sufficient large value to cause droplet fission (Fig. S5). \red{The steady-state lengthscale $\xi_{steady}$, was maintained by the balance of droplet fission and fusion events, where $\xi_{steady}$ was comparable to the inverse of the average interface curvature $\kappa_{steady}$.} Concomitantly with the plateauing of $\xi$,  active flow speed became constant (Fig. 1F). These results demonstrate activity-suppressed coarsening dynamics, which created an emulsion-like state wherein finite-sized droplets continuously merge, break apart and exchange their content (\red{Fig. 1G}, Movie S3, S4). \red{The volume fraction of the active and passive phases were nearly equal (Fig. S2). Low volume fraction of active fluid generated \MCM{similar} steady states.} Finite-sized domains are reminiscent of theoretical prediction in motility-induced-phase-separation of isotropic active particles~\cite{Singh2019}. However, in contrast to theory, the active fluid in our experiments is anisotropic and perturbs an underlying equilibrium phase separation. 

To gain insight into how active stresses drive interfacial fluctuations we formed a macroscopic interface by gravity-induced bulk phase separation (Fig\red{s}. 2A\red{, S6}). In equilibrium, molecular motion works against the density difference $\Delta\rho$ and interfacial tension $\gamma$ to roughen the liquid-liquid interface. Typical disturbances of PEG-dextran interfaces, bereft of activity, are $\sim 100$ nm in amplitude, resulting in boundaries that appear flat when viewed with our imaging setup (Figs. 2A\red{,} S7). When driven out of equilibrium, however, interfaces exhibited giant undulations that were visible by naked eye (Movie S5). As motor concentration increased, interfaces became multivalued with frequent overhangs, indicating that active stresses directly control interface configurations (Fig. 2A, Movie S6).

The interplay of activity and capillarity is clarified by measuring the interfacial fluctuation spectrum. To this end, local interface tangent angles $\theta (s,t)$ were sampled at a time $t$ as a function of the arc-length distance $s$ along the interface (Fig. 2A). Interfacial fluctuations were described by time-averaged power spectra $S(k)=\langle\|\theta_k\|^2 \rangle_t$, with $\theta_k=\int\!\mathrm{d}s\, \theta(s,t) e^{-iks}$. Due to equipartition of thermal energy among Fourier modes, the spectrum of equilibrium interfaces is $S(k)\sim Tk^2/(k^2 + k_c^2)$, where $T$ denotes temperature . The capillary wave-number $k_c=\sqrt{\Delta\rho g/\gamma}$, sets a crossover from a gravity dominated regime at large scales $S(k)\sim k^2$ to a plateau  at small scales where surface tension attenuates fluctuations. Active interfacial fluctuations were markedly different. Active spectrum $S(k)$ increased for small wave-numbers (Figs. 2B, S7\red{,8}). After reaching a maximum for $k_m\sim30\, mm^{-1}$, it decayed as $S(k)\sim k^{-3}$, instead of plateauing as in equilibrium. While the shape of $S(k)$ remained the same for all KSA concentrations, the root-mean-square tangent angle $\theta_{rms}$ increased linearly with activity (Fig. 2B, inset). Using the crossover at $k_m$ as a determinant of the fluctuation amplitude, it would take an effective temperature of $\sim 10^{11} K$ to achieve equilibrium interfaces whose roughness is comparable to those measured at the lowest activities. 

The  dynamics of activity-driven interfacial fluctuations exhibit non-trivial spatiotemporal correlations. To gather sufficient statistics, we imaged $\sim$10 mm-long active interfaces over a $2$ hour interval. Space-time maps of local interface height $h(x,t)$ exhibited diagonally streaked crests and troughs that were suggestive of propagating waves (Fig. \red{2C}). These translational modes were also evident in time-lapse movies (Movie S6). To characterize these modes, we measured the dynamic structure factor (DSF) of the interface height $D(k,\omega) = \\ \int\!\mathrm{d}x\mathrm{d}t \, e^{ik  x+ i\omega  t} \langle h(x',t')h(x'+x,t'+t)\rangle_{x',t'}$ (Fig. \red{2D}). Over a finite range of wave numbers, the DSF exhibited peaks at finite frequencies $\omega_p$, confirming the presence of propagating modes (Fig. \red{2E}). Increased KSA concentration resulted in higher $\omega_p$ for the same wave-numbers; thus, activity controled the phase velocity (Fig. \red{2F}). 

The giant non-equilibrium fluctuations and propagating wave modes result from the interaction of active flows in the bulk dextran phase with interfacial elasticity. To elucidate the processes driving active interfaces, we numerically integrated 2D hydrodynamic equations that describe a bulk-phase-separated active fluid~\cite{Giomi2014,Blow2014}. The two coexisting phases were modelled as \red{incompressible} Newtonian fluids under gravity that experience confinement friction in the low Reynolds number limit~\cite{SI}. The top phase was passive, while the \red{velocity of the} bottom phase \red{$\textbf{v}$} was \red{governed by
\begin{equation}
    \gamma_v \textbf{v} -\eta \nabla^2 \textbf{v}  = -\boldsymbol{\nabla} P +\boldsymbol{\nabla}\cdot\boldsymbol{\sigma}\;,
\end{equation}
with $P$ the pressure, $\eta$  the viscosity, and $\gamma_v$  the confinement friction. The stresses $\boldsymbol{\sigma}$ driving the flows were assumed to be generated by an active liquid crystal producing extensile active stresses, $\boldsymbol{\sigma}^a = \alpha \textbf{Q}$, with $\alpha<0$.}. The local orientational order was quantified by a traceless tensor $Q_{ij}=\langle \hat{n}_i\hat{n}_j-\delta_{ij}/2\rangle$  averaged over molecular orientations $\hat{\textbf{n}}$.   Active shear flows engendered orientational order via flow-induced alignment.  These assumptions are summarized in the continuum equation
\begin{eqnarray}
%&\sigma_{ij} = \sigma^a_{ij} +\sigma^e_{ij}\;,\nonumber\\
\frac{D \textbf{Q}}{D t}+ \left[\boldsymbol{\omega},\textbf{Q}\right] = \lambda \textbf{u} +\frac{1}{\gamma_Q}\textbf{H}\;,
\end{eqnarray}
 with $\boldsymbol{\omega}$  the vorticity tensor, $\textbf{u}$  the strain rate and $\lambda$ the flow alignment parameter. 
 %$\sigma^e_{ij}$ and 
 $\textbf{H}$ denotes elastic 
 %stresses and conjugate 
 forces that arise from the liquid crystal free energy, and $\gamma_Q$ the rotational viscosity of microtubule bundles~\cite{SI}. In the absence of activity the liquid crystal is in the isotropic phase, consistent with the microtubule density used in experiments.

The hydrodynamic model reproduces key experimental observations. Finite-sized chaotic flows, driven by active stresses, continuously deform the liquid-liquid interface (Fig. \red{3A}, Movie S7). Similarly to experiments, the interfacial power spectra showed a crossover between growth at small wave-numbers and decay at large wave-numbers, while the root-mean-square tangent angle increased linearly with $|\alpha|$  (Fig. \red{3B}). The numerically obtained DSF \red{also} exhibited signatures of active travelling waves (Fig. \red{3C-E})\red{, as in the experiment}. The wave frequencies $\omega_p(k)$ increased with activity, showing active-stress-dependent wave velocity. Our numerical model also suggests a non-inertial mechanism of active waves~\cite{Soni2019}, which differs from conventional \red{inertia-dominated} capillary waves~\cite{Langevin1992,Aarts2004}. In the context of our hydrodynamic model, the active waves arise from the coupling between the interface vertical displacement $h$ and orientational order $Q$ in the interfacial region~\cite{Blow2014}. Stress balance at the interface predicts that the orientational order drives interfacial deformation as $\partial_t h \sim v^a_\perp\sim \alpha Q$, where $v^a_\perp$ is the active contribution to the flow velocity normal to the interface. In turn, passive flows $v^p_\perp$  relaxing the interface at a wave-number dependent rate $\nu(k)$ feed back to induce local liquid crystalline order $\partial_t Q\sim \lambda ik v^p_\perp\sim -\nu(k) h$. Consequently, interface height obeys a wave equation $\partial_t^2 h\sim -\alpha \nu(k) h$~\cite{SI}. Accordingly, travelling wave velocities increase with active stress, which is in agreement with both experiments and numerics. 

Propagating waves might be a generic feature of active boundaries~\cite{Soni2019, Soni2019a}. More broadly, the active-stress-dependent wave dispersion mirror those of  elastic waves in entangled polymers solutions~\cite{Varshney2019}. While the numerical hydrodynamics reproduced qualitative features of the experimentally observed active fluctuations and waves, there were important quantitative differences. In particular, with increasing activity, numerics predict increase in both interface fluctuations and bulk velocity. In contrast, in experiments active flows remain constant between 200 and 350 nM KSA, while interfacial fluctuations increase (Fig. S9). Moreover, in numerics, the maximum wave-number $k_m$ increased with activity, while remaining constant in experiments (Figs. 2B, 3B, S10).

To demonstrate the unique properties of active interfaces, we studied their structure next to a solid boundary (Fig. \red{4}A\red{,B}). In the absence of motors, the interface assumed an exponential profile $h\propto e^{-x/l_e}$ with a decay length of $\ell_e\sim 45 ~\mathrm{\mu m}$, \red{which we identified with}  the capillary length  $\ell_c=\sqrt{\gamma/\Delta \rho g}$. At the wall, the rise in the interface height was $\sim 70~ \mathrm{\mu m} $, which is close to the maximum capillary rise of $\sqrt{2} \ell_c$, indicating complete wetting~\cite{SI}. At intermediate KSA concentrations, the capillary rise exhibited active fluctuations around the equilibrium exponential height profile (Fig. \red{4C}), and the time-averaged center-of-mass height of the wetting region increased slowly (Fig. 4C). Above a critical value of 300 nM KSA, activity generated a new interfacial structure. Specifically, we observed formation of a  $\sim$20 $\mathrm{\mu m}$ thick dynamical wetting layer, which climbed several hundred microns above the equilibrium capillary rise (Fig. \red{4}A, Movie S8). \red{Within this layer, microtubule bundles preferentially aligned with the wall (Fig. 4B).} Coinciding with the appearance of the microtubule-rich wetting layer, the capillary rise sharply increased (Fig. 4C). These observations demonstrate an activity driven wetting transition beyond the complete wetting of a passive fluid. 

We performed numerical simulations of the active-interface adjacent to a vertical boundary~\cite{SI}. The liquid crystal director was anchored parallel to the wall, and the surface-liquid energy $\gamma_w$ corresponds to an equilibrium contact angle 10$^\circ$ (Fig. \red{4D}). Similar to experiments, the average height profile had an exponential decay (Fig. \red{4E}, inset). As the activity $\alpha$ increased from zero, the height of the contact point increased. Furthermore, above $|\alpha| =  10$ mPa, the active fluid generated a thin wall-adjacent layer, indicating a transition from partial to complete wetting (Fig. \red{4D}, Movie S9). The capillary rise was supported by a $\ell_w\sim 3 \,\mu\mathrm{m}$ thick, vertically aligned liquid crystalline domain with $Q\sim 1$. This domain generated coherent active stress along the wall $\sigma^a =\alpha$, which supported the interface rise. Balancing the active tension $\yzh{\gamma_a\equiv}|\sigma^a| \ell_w$ at the contact point with \red{wall adhesion $\gamma_w$, surface tension $\gamma\cos\theta_a$, and gravity $F_g$} resulted in a boundary condition for the climbing interface (Fig. S1\red{1})~\cite{SI},
\begin{equation}
    |\sigma^a| \ell_{w} + \gamma_{w} =\gamma \cos\theta_a + F_g\,.
\end{equation}
Predictions of the center-of-mass height of the capillary rise, using Eq. (3), show a crossover from slow to fast growth at $\alpha=10$~mPa which is in agreement with the partial to complete wetting transition seen in simulations (Figs. 4E, S1\red{2}). 

\red{Active interfaces provide a unique experimental probe to estimate the magnitude of the active stress, a critical parameter that governs dynamics of active fluids. To avoid resorting to various assumptions on the numerical model, we analytically solved Eq. (1) \MCM{treating the stress $\sigma$ as a random field with correlations} $\langle \sigma_{ij} (\mathbf{r},t)\sigma_{ij}(\mathbf{0},0)\rangle = \sigma_{rms}^2 e^{-|\mathbf{r}|/\ell_a-|t|/\tau_a}$, where correlation length $\ell_a$ and time $\tau_a$ are \MCM{identified with} those of the \MCM{bulk} active flow (Fig. S13)~\cite{SI}. The analytical model captured the interface fluctuations spectrum $S(k)$ without fitting parameters, revealing that its non-monotonic shape resulted from active flows with scale-dependent kinetic energy spectrum (Figs. 5A, S10, Eq. (S32))~\cite{SI,Martinez-Prat2021}. In contrast, fluctuations of equilibrium interfaces are driven by thermal energy $\sim k_BT$ that is equally distributed among all scales~\cite{SI}. By integrating $S(k)$ over all wave numbers, the active stress is predicted to increase proportionally to tangent angle fluctuations $\sigma_{rms} = p \theta_{rms}$, with $p \simeq 9$ mPa/rad~(Fig. 5B, Eq. (S35)).}

\red{To independently verify these numbers we note that the force balance Eq. (\red{3}), associated with active wetting, provides an alternative method of estimating active stress. For intermediate KSA values, prior to the appearance of the active wetting layer, the active stress \MCM{estimated from wetting} is comparable to those extracted from interface fluctuations (Fig. 5B). Above 300 nM KSA, active stresses are a few times larger than those obtained from interface fluctuations. These large values might be a consequence of flow-enhanced alignment of microtubule bundles within the thin wetting layer. The formation of the active wetting layer at finite activity, however, is outside the scope of the static stress balance embodied in Eq. (3). A more complete description of the wetting transition would include dynamical considerations, such as active wave propagation and gravitational sedimentation, that appear to be essential for the formation and turnover of the wetting layer.}

In summary, we demonstrated a rich interplay between active fluids and soft deformable interfaces. On the one hand, liquid interfaces provide a quantitative probe that can reveal intrinsic properties of the active fluids, such as active stress. On the other hand, bulk active fluids drive the extreme interfacial deformations that yield intriguing non-equilibrium dynamics, including arrested phase separation, stress-dependent non-inertial propagating waves and activity-controlled wetting transitions. Our findings provide a promising experimental platform to design shape-changing adaptable soft materials and machines whose capabilities begin to match those observed in biology~\cite{Joanny2012,Tjhung2015,Young2021}.

\bibliography{Active_interfaces}
\bibliographystyle{unsrt}
\clearpage

\subsection*{Acknowledgments}
We thank Fernando Caballero for enlightening discussions. \textbf{Funding}: Experimental work was supported by the U.S. Department of Energy, Office of Basic Energy Sciences, through award DE-SC0019733 (R.A., I.K., and Z.D.). Theoretical analysis was primarily supported by NSF-DMR-1720256 (iSuperSeed) with additional support from NSF-DMR-2041459 (Z.Y. and M.C.M.). I.K. acknowledges support from the HFSP  cross-disciplinary fellowship LT001065/2017-C. R.A. acknowledges support from NSF-GRFP-1650114. We also acknowledge the Brandeis biosynthesis facility, supported by (MRSEC) grant DMR-2011846.
\textbf{Author contributions}: R.A., S.W. and I.K. conducted experimental research;  I.K., S.W., R.A. and Z.Y. analyzed experimental and theoretical data; Z.Y. applied finite volume OpenFOAM solver code to integrate continuum equations; I.K., Z.Y. and M.C.M. conducted theoretical modeling and interpretation of data. I.K., M.C.M. and Z.D conceived the work. I.K., M.C.M. and Z.D. wrote the manuscript. All authors reviewed the manuscript. \textbf{Competing interests}: The authors declare no competing interests.
\textbf{Data availability}: Data for interfacial height and fluid velocity as well as executable
code for numerical simulations are available at \cite{dryad}. All other data
are reported in the main text and  supplementary information. 

\subsection*{Supplementary Materials}

Materials and Methods\\
Figs. S1 to S13\\
Movies S1 to S9\\
References \cite{Ward2015,Squires2009,Gidi2018,Sanchez2013,Tayar2022,Ciomaga2017,Thielicke2021,Hirt1981,Brackbill1992,Srivastava2016,Santhosh2020,DeGennes1993,Moukalled2016,Deshpande2012,Safran2003,Grant1983,Harden1991,Flekkoy1995,Flekkoy1996,Bray2001,Doostmohammadi2018,Gagnon2020,Kimura1985,Blow2017,Coelho2021,Berg2010}

\begin{figure}
\centering
\includegraphics[scale=1]{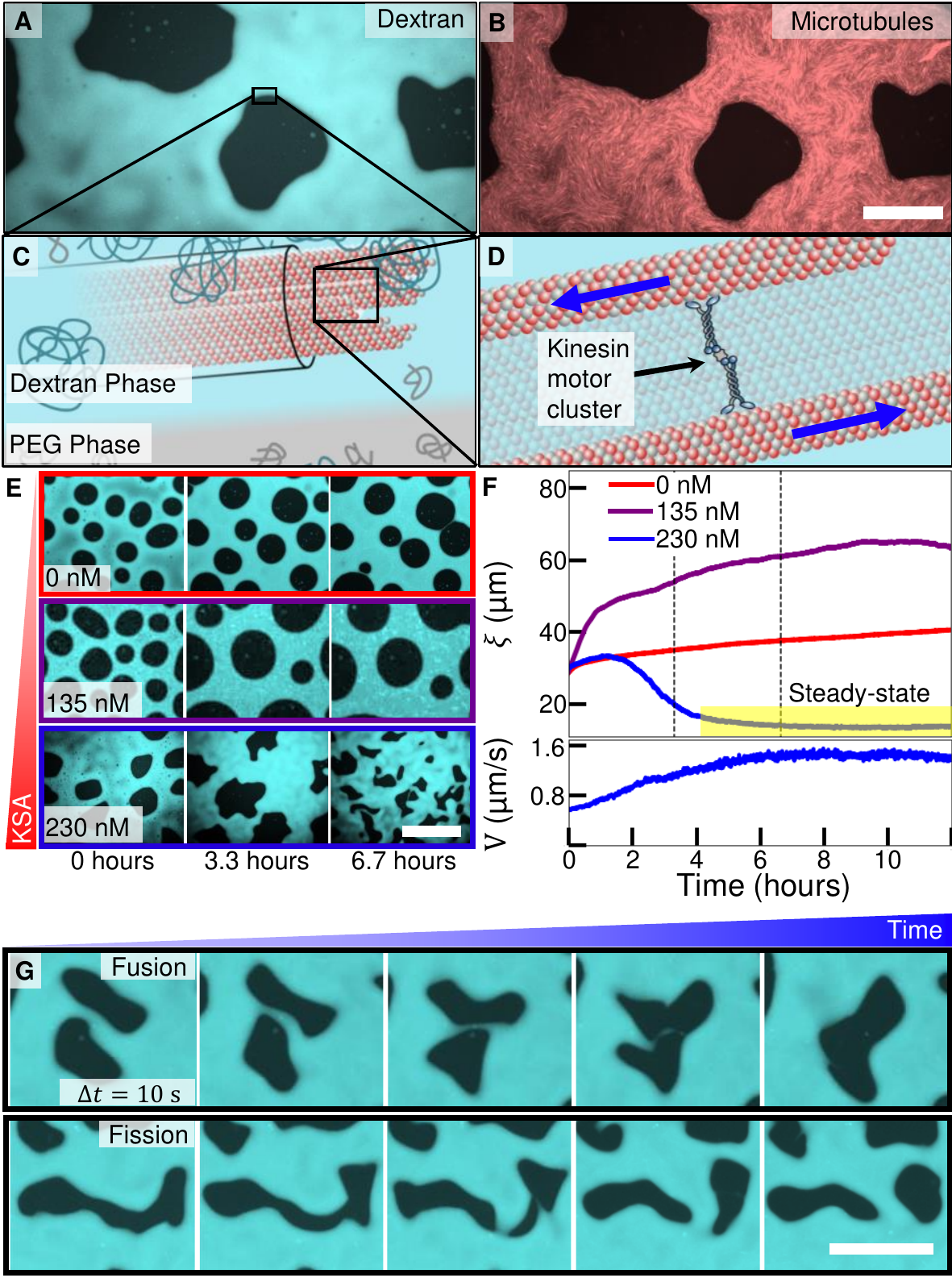}
\caption{\textbf{Active liquid-liquid phase separation} (\textbf{A}) Coexisting PEG (dark) and dextran (cyan) rich domains.  (\textbf{B})  Labeled microtubules (red) are dissolved in the dextran phase. Scale bar, 75 $\mathrm{\mu m}$. (\textbf{C}) Microscopic-scale depiction of phase separation. Minority PEG polymers (gray) in the dextran-rich phase induce microtubule bundling.  (\textbf{D}) Kinesin clusters drive interfilament sliding. (\textbf{E}) Time evolution of the active LLPS at three KSA concentrations. Scale bar, 350 µm.  (\textbf{F}) Top: Correlation length evolution $\xi(t)$ for three KSA concentrations. For 230 nM KSA, $\xi$ plateaus at long time (yellow highlight). Bottom: Root-mean-square velocity of turbulent flows in the dextran phase at 230 nM KSA. (\textbf{G}) Fusion and fission of PEG droplets. Sample chamber thickness, 30 $\mathrm{\mu m}$. Scale bar, 100 $\mathrm{\mu m}$.}
\end{figure}

%\begin{figure}
%\centering
%\includegraphics[scale=1]{main_text_figures/Figure2_v10.pdf}
%\caption{\textbf{Active interfacial fluctuation power spectra.} (\textbf{A}) Conformations of bulk-phase-separated interfaces with increasing activity (KSA) concentration. $\theta (s)$ defines the local interface tangent angle. (\textbf{B}) Tangent angle power spectra obtained by averaging over $\sim$3 mm interface length, for $\sim$2 hours, 6 hours after sample preparation. Inset: Root-mean-square tangent angle  as a function of KSA concentration. Chamber thickness, 60 $\mathrm{\mu m}$. Scale bar, 150 $\mathrm{\mu m}$.}
%\end{figure}

\begin{figure}
\centering
\includegraphics[scale=1]{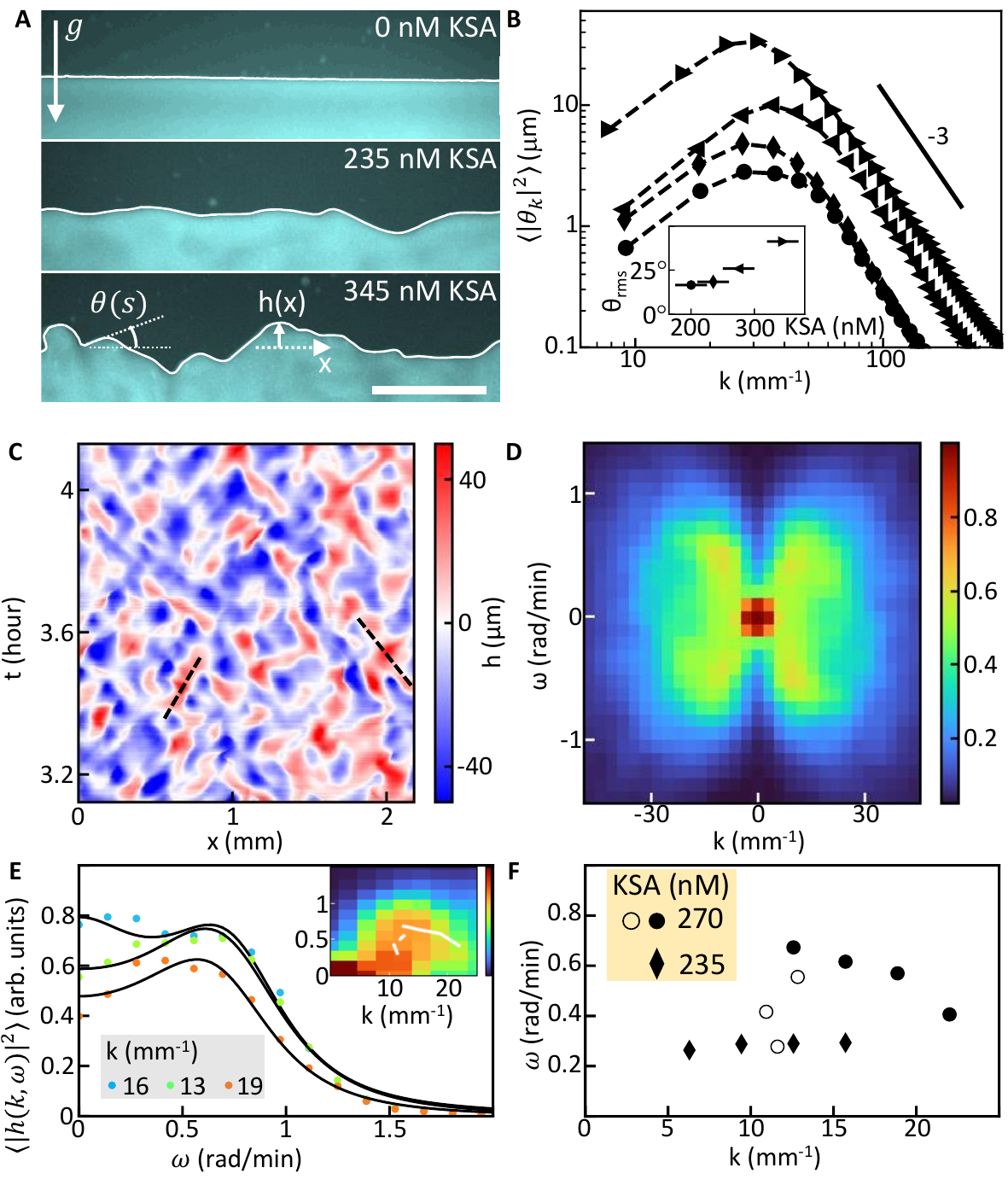}
\caption{\textbf{Active propagating waves.} (\textbf{A}) Conformations of bulk-phase-separated interfaces with increasing motor (KSA) concentration. $\theta (s)$ defines the local interface tangent angle as a function of arclength. $h(x)$ is the local interface height. Chamber thickness, 60 $\mathrm{\mu m}$. Scale bar, 150 $\mathrm{\mu m}$. (\textbf{B}) $\theta$ power spectra obtained by averaging over $\sim$3 mm interface length, 6-8 hours after sample preparation. Inset: Root-mean-square $\theta$  vs. KSA concentration.  (\textbf{C}) Space-time map of $h$. Disturbances propagating along the interface (dashed lines). Interfaces were corrected for drift and tilt.   (\textbf{D}) The square-root DSF averaged 4-6 hours after sample preparation. (\textbf{E}) DSF sections at constant wave-number (colored dots). Black lines are best approximations with $F(\omega) = a\left((\omega/\omega_1)^2+1\right)^{-1}+b\left[\left((\omega/\omega_0)^2-1\right)^2+(\omega \Delta \omega/\omega_0^2)^2\right]^{-1}$, where $a,b,\omega_1,\omega_0,\Delta\omega$ are adjustable parameters. Data taken over 2-4 hours after sample preparation. Inset: Frequency peaks $\omega_p = \sqrt{\omega_0^2-\Delta\omega^2/2}$ overlayed on DSF intensity (full line). DSF maxima for constant $\omega$ (dashed line). For (\textbf{C}-\textbf{E}), KSA concentration, 270 nM. (\textbf{F})  Peak frequencies of the propagating modes $\omega_p$ (full symbols). DSF maxima at constant frequency sections (empty symbols).}
\end{figure}

\begin{figure}
\centering
\includegraphics[scale=1]{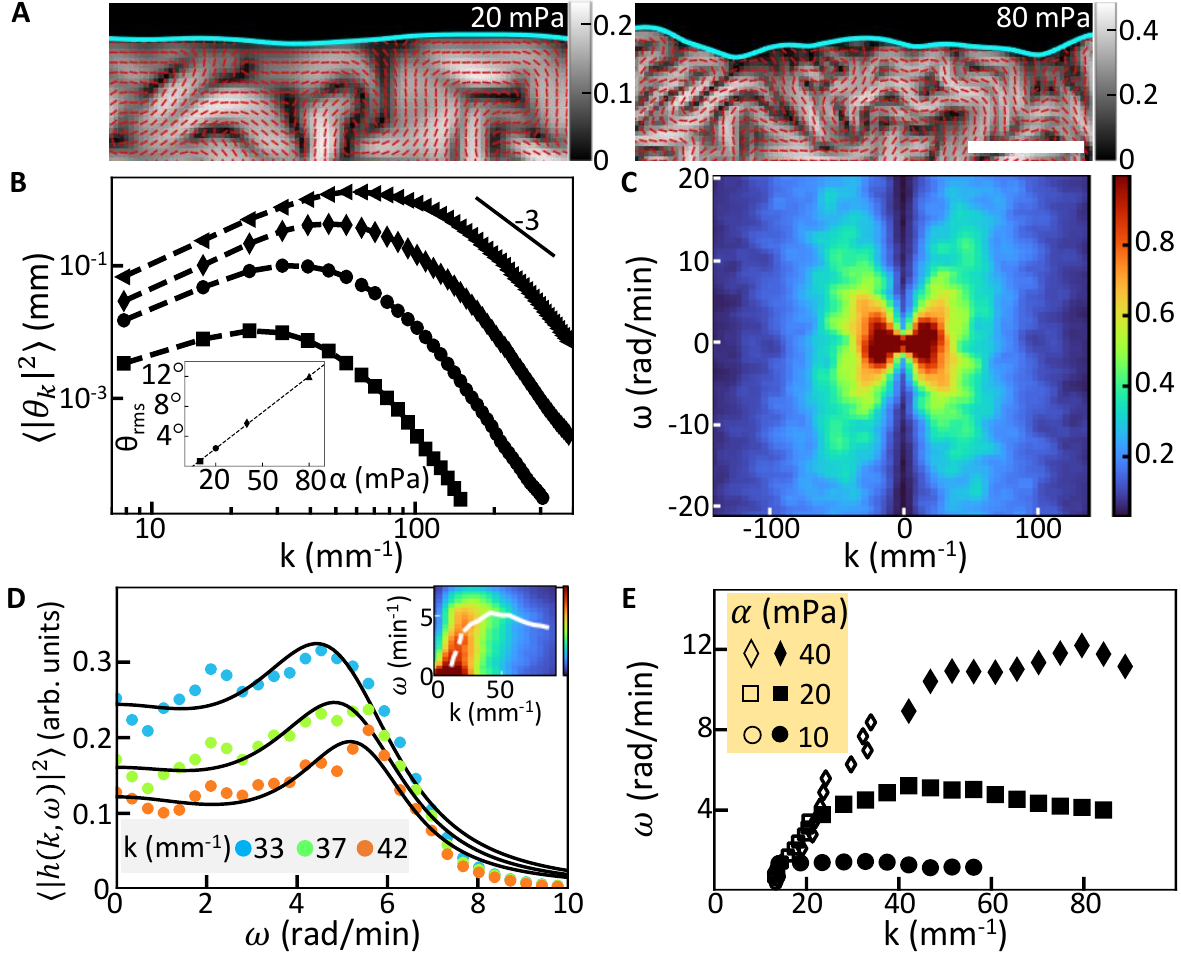}
\caption{\textbf{Numerical hydrodynamics of active interfaces.} (\textbf{A}) Phase boundary (cyan) separating a passive fluid (black) and an active isotropic liquid crystal. The latter depicted with local order parameter (grayscale) and orientation (red lines). \red{Legends denote activities $|\alpha|$}. Scale bar, 100 $\mu$m. (\textbf{B})  $\theta$ power spectra. Simulation interval, 2 $\mathrm{mm}$. Correlation maximum lag distance, 400 $\mu$m. Inset: Root-mean-square $\theta$ vs. $|\alpha|$. (\textbf{C}) Square root DSF. $\red{|\alpha|} = 40 \, \mathrm{mPa}$. Maximum lag time, 270 sec. Maximum lag distance, 670 $\mu m$. Total simulation time, $3$ hours. Simulation interval, 2 $\mathrm{mm}$. (\textbf{D}) Constant wave-number sections of  DSF intensity(filled circles).  Best approximations $F(\omega)$, as in Fig. \red{2E} (black lines). \red{$|\alpha|=20$ mPa.} Inset:  $\omega_p(k)$ overlay on DSF intensity (full line). DSF maxima for constant $\omega$ (dashed line). (\textbf{E}) Peak frequencies of the propagating modes $\omega_p$ (full symbols). DSF maxima for constant $\omega$ (empty symbols). Activities are noted in absolute values.}
\end{figure}

\begin{figure}
\centering
\includegraphics[scale=1]{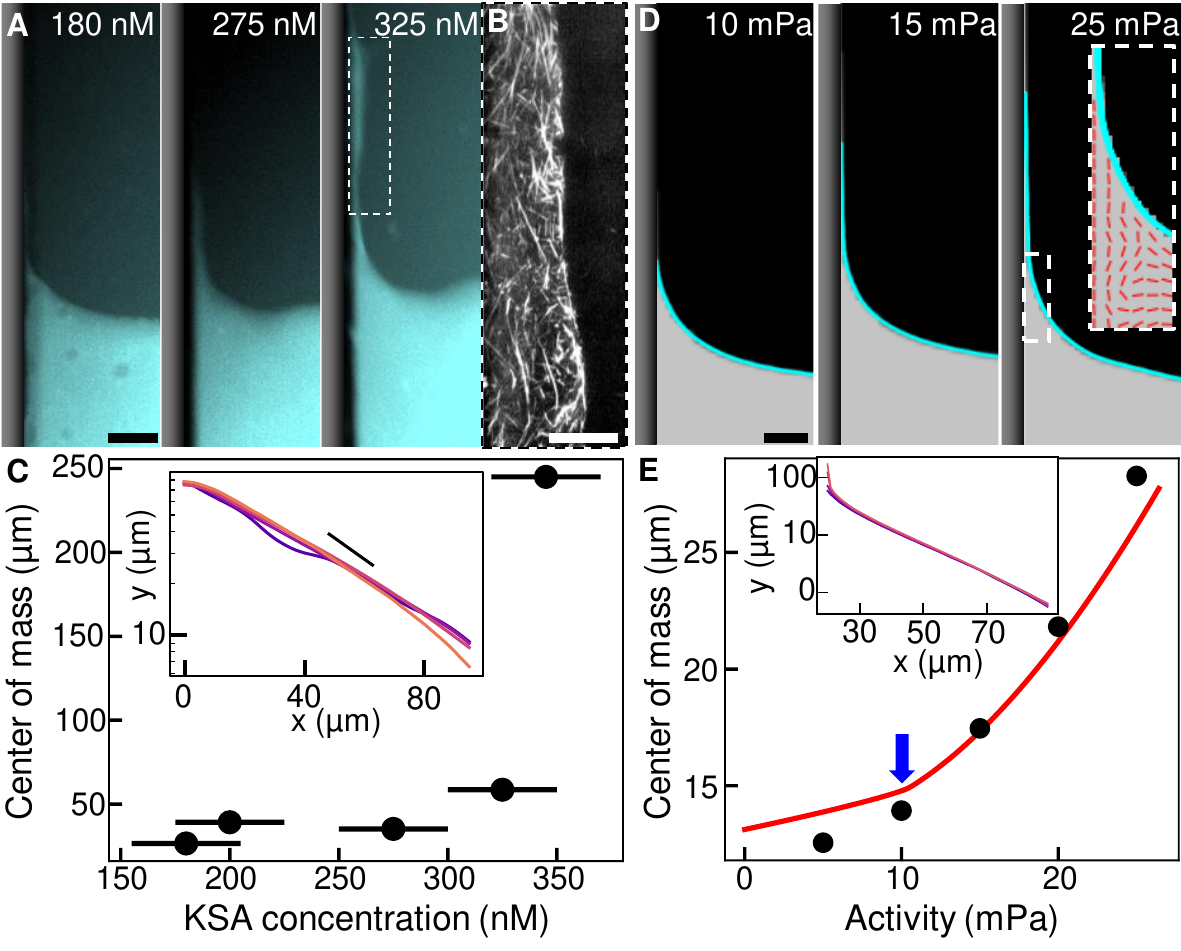}
\caption{\textbf{Active wetting transition.} (\textbf{A}) Active fluid wetting a polyacrylamide coated vertical boundary at three KSA concentrations. Scale bar, 50 $\mathrm{\mu m}$. (\textbf{B}) Magnified section of the wetting layer at 345 nM KSA showing local orientations of microtubule bundles. Scale bar, 20~$\mu$m. (\textbf{C}) Average center-of-mass height of active fluid within 5$l_e$ of the vertical boundary where $l_e = 45 \, \mathrm{\mu m}$. Zero height defined as the average bulk interface position. Each point is the mean of two experiments. Uncertainty in sample preparation (horizontal bars).  Inset: Average wetting height profiles as a function of distance from the vertical boundary. KSA concentrations, 180 (blue) to 270 (red) nM. Black line  $\sim e^{-x/l_e}$ . (\textbf{D}) Wetting profiles from numerical simulations. Equilibrium contact angle, $\theta_e=10^\circ$. Interface position (cyan).  Scale bar, 50~$\mu$m. Inset: Local liquid crystalline orientation (red lines). (\textbf{E}) Center-of-mass height of the numerical wetting profile. Red line, prediction of Eq. (3) with $l_w = 2.5\, \mathrm{\mu m}$. Onset of complete wetting (blue arrow). Inset: Average numerical wetting profiles. Activities, 10 (blue) to 40 (red) mPa. }
\end{figure}

\begin{figure}
\centering
\includegraphics[scale=1]{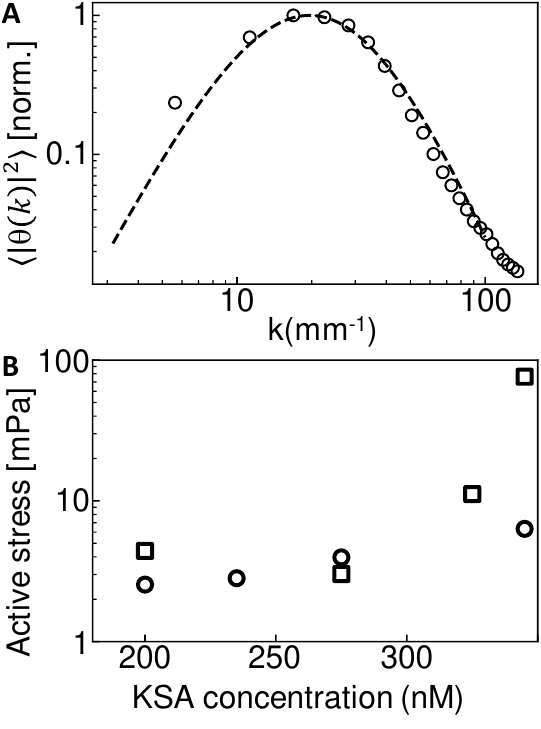}
\caption{\textbf{Active stress estimation from interface fluctuations and wetting.} (\textbf{A}) Normalized tangent angle power spectrum. KSA concentration, 200 nM (circles). Analytical hydrodynamic prediction (Eq. (S35), dashed line). (\textbf{B}) Root-mean-square active stress estimates from interfacial fluctuations $\sigma_{rms} = p\theta_{rms}$ (circles). Experimental $\theta_{rms}$ (Fig. 2B, inset). Analytical hydrodynamic theory (Eqs. (1,S35)) predicts $p$ = 8.8 mPa/rad. Bulk flow correlations yield active length scale $\ell_a = 65~ \mu$m and time scale $\tau_a = 77$ sec (Fig. S13). Active stress estimates from wetting $\sigma^a = 4(h_{cm} - h_{cm}^0)\Delta\rho g$ (squares). $h_{cm}$ is the wetting center-of-mass height, and $h_{cm}^0$ equals $h_{cm}$ at 180 nM KSA. (Fig. 4C). }
\end{figure}

\clearpage

\beginsupplement
\begin{titlepage}
\begin{center}

\Large Supplementary Materials for
\bigskip

\large Dynamics of Active Liquid Interfaces
\bigskip

\normalsize Raymond Adkins, Itamar Kolvin, Zhihong You, Sven Witthaus,
M. Cristina Marchetti and Zvonimir Dogic
\bigskip

\end{center}

\bigskip
\bigskip

\end{titlepage}

\phantom{References \cite{Marchetti2013,Giomi2014,Blow2014,Martinez-Prat2021,Ward2015,Squires2009,Gidi2018,Sanchez2013,Ciomaga2017,Thielicke2021,Brackbill1992,Srivastava2016,Santhosh2020,DeGennes1993,Hirt1981,Moukalled2016,Deshpande2012,Safran2003,Grant1983,Harden1991,Flekkoy1995,Flekkoy1996,Bray2001,Doostmohammadi2018,Gagnon2020,Kimura1985,Blow2017,Coelho2021,Berg2010,Tayar2022}}

\renewcommand{\theequation}{{S\arabic{equation}}}
\setcounter{equation}{0}
\noindent\textbf{Experimental methods}\\

\noindent\underline{Dextran fractional precipitation.} 

Control of LLPS required low-polydispersity, high-molecular-weight dextran. Dextran (1.5-2.8 MDa, Sigma-Aldrich) was separated into fractions of narrow molecular weight distributions via ethanol precipitation. Ethanol was gradually added to a solution of 0.2\% dextran under vigorous stirring at 23 $^\circ$C. After reaching 31\% (w/w) ethanol, precipitates were removed by centrifugation (20 min at 17,000g, 
Fiberlite F9-6 x 1000 LEX fixed angle rotor, Thermo Scientific). Ethanol was then added to a concentration of 32\% (w/w). The precipitate was collected by centrifugation and resuspended in water. Solvent was removed via lyophilization, and the powder was reconstituted in water to 20\% (w/w) dextran. This stock solution was used to make LLPS solutions.  

\noindent\underline{PEG-dextran LLPS.} 

Polymers of high molecular weight chosen to create mixtures that are both active and phase separated. Inter-microtubule sliding occurs in a finite polymer concentration range, where depletion forces are sufficiently strong to induce microtubule bundling without friction~\cite{Ward2015}. Mixtures comprising 2.38\% (w/w) fractionated dextran and 1.55\% (w/w) poly(ethylene glycol) (PEG) (35 kDa, EMD Millipore), reconstituted in M2B buffer (80mM K-pipes, 2mM MgCl\textsubscript{2}, 1mM EGTA, pH 6.8), enabled both phase separation and motor-driven inter-filament sliding. To distinguish the two phases, 2,000 kDa amino-Dextran (Fina Biosolutions) was labeled with Alexa-Fluor 488 NHS Ester  (ThermoFisher Scientific) and was added at a final concentration of $<$ 0.1\% (w/w). 

To characterize the LLPS, we let mixtures completely phase separate under gravity for one day. Top (PEG-rich) and bottom (dextran-rich) phases were extracted, and their densities were measured using a density meter (DMA 4100, Anton-Paar). The densities of the dextran and the PEG phases were $\rho_P = 1.0151 g/mL$ and $\rho_D = 1.024g/mL$ respectively. The viscosity of each phase was determined by microrheology \cite{Squires2009}. The viscosities of the PEG-rich and the dextran-rich phases were $\eta_P = 5 \mathrm{mPa\, s}$ and $\eta_D = 25 \mathrm{mPa\, s}$ respectively. \red{Interfacial tension between the two phases was determined from the exponential decay length of the capillary rise next to a polyacrylamide coated wall. At low KSA concentrations, the decay length was $l_e = 45 \mu$m (Fig 5C, inset). The relation $l_e = l_c = \sqrt{\gamma/\Delta\rho g}$, where the density difference is $\Delta\rho = \rho_D-\rho_P$, and $g$ is the acceleration of gravity, yields the interfacial tension $\gamma = 0.18 \mu$N/m.}
%Interfacial tension between the two phases was determined by analyzing droplet coalescence dynamics~\cite{Jeon2018}. Neighboring dextran droplets, sitting on an oil-water interface, came in contact with each other due to thermal motions. Two fusing droplets relaxed $A(t)\sim e^{-t/\tau}$ from an ellipsoidal to a spherical shape, quantified by the geometric ratio $A = (L-W)/(L+W)$ where $W$ and $L$ are the dynamic length and width of the droplet. The interfacial tension is related to the equilibrium radius $R$ and relaxation time $\tau$ via $\gamma = \frac{(2\lambda+3)(19\lambda+16)}{40(\lambda+1 )}  \eta_P R/\tau$, with $\lambda=\eta_D/\eta_P$. Using this expression, the equilibrium PEG/dextran interfacial tension, without microtubules or kinesin motors, was $\gamma = 0.8\pm 0.2\, \mathrm{\mu m }$.

\noindent\underline{Chamber preparation.} 

Chambers were constructed of glass slides that were coated with a PEG brush (mPEG 5k-silane, BiochemPEG)~\cite{Gidi2018}. PEG-coating resulted in preferential wetting by the passive PEG phase. Achieving a uniform coating was essential to prevent pinning of the dextran phase to the chamber walls. Glass was cleaned by sonicating in 1\% Hellmanex, and then etched in 0.5 M NaOH for 30 minutes. Slides were dried at 90 $^\circ C $ in the presence of a desiccant for 10 minutes. Silanized PEG, reconstituted in anhydrous DMSO to a concentration of 5\%, was sandwiched between glass slides and left to react for 30 minutes at 90  $^\circ C$. Slides were then rinsed in water and dried with a nitrogen stream. Chambers were assembled immediately. For wetting experiments, \#0 coverslips were coated with polyacrylamide according to established protocol \cite{Sanchez2013}. Chambers were constructed by sandwiching two polyacrylamide coated coverslips between PEG-coated slides.

\noindent\underline{PEG-dextran active-LLPS.} 

We assembled active-LLPS by adding the phase-separating polymers to microtubules and clusters of kinesin motors. Kinesin K401-streptavidin (KSA) motor clusters and GMPCPP-stabilized microtubules were purified and prepared as described previously~\cite{Tayar2022}. The active mixture was prepared in M2B buffer containing antioxidants (2 mM Trolox, 3.3 mg/mL glucose, 5 mM DTT, 200 $\mu$g/mL glucose oxidase and 35 $\mu$g/mL catalase) to reduce photobleaching, ATP (1420 uM), an ATP regeneration system (26 mM phosphoenol pyruvate (Beantown Chemical, 129745) and 2.8\% (v/v) pyruvate kinase/lactate dehydrogenase enzymes (Sigma, P-0294)). We added microtubules to a final concentration of 0.67 mg/mL and KSA at variable concentration.

\noindent\underline{Correlation length evolution in phase separation experiments.}

To quantify the phase separation dynamics, images of dextran fluorescence were thresholded at each time point to produce a binary intensity map $I(\vec{r},t)$, where $I=1$ for dextran-rich domains and $I=-1$ for PEG-rich domains, and the radius $\vec{r} = (x,y)$ (Fig. \ref{fig:correlation}A,B). The two-point correlation function $C(\Delta \vec{r},t) =\langle I(\vec{r}+\Delta\vec{r},t)I(\vec{r},t)\rangle$ was azimuthally averaged to produce the radial correlation function $C(r,t)$. The correlation length $\xi(t)$ was defined such that $C(r=\xi(t),t)=0.5$ (Fig. \ref{fig:correlation}C,D), and its evolution was tracked (Fig. \ref{fig:correlation}E).

\noindent\underline{Evolution of correlation length and interface curvature at 230nM KSA.}

In samples approaching a steady state, increase in the interface curvature preceded the decay of the correlation length. Local interface curvatures were computed as 
\begin{equation}
    \kappa = \frac{f_{xx}f_y^2-2f_{xy}f_xf_y+f_{yy}f_x^2}{(f_x^2+f_y^2)^{3/2}}
\end{equation}
where $f(x,y)$ is the dextran fluorescence intensity at pixel $(x,y)$ and subscripts denote partial derivatives \cite{Ciomaga2017}. $\kappa$ was averaged over all interfaces in the field of view at each time point.
Initially, the average curvature increased before the correlation length started to decay (Fig. \ref{fig:correlation_and_curvature}).  The correlation length evolution lagged by 30 min behind the average interface curvature for the first 1.5 hours of the experiment. After 2 hours, both correlation length and inverse curvature evolved synchronously as the system approached steady state.

\noindent\underline{Detection of bulk-phase-separated interfaces.}

Interfaces were detected using a multi-step procedure:

(1) Dextran fluorescence images were divided by a background image, thresholded, and numerically differentiated to extract domain edges. Edges that straddle the image horizontal axis were selected, skeletonized, and pruned to produce an initial contour at single pixel resolution (Fig. \ref{fig:InterfaceDetection}A-C).

(2) For each point on the contour $(x, y)$, image intensity is interpolated at sub-pixel resolution along the local interface normal $(n_x, n_y)$ in a 5x5 pixel neighborhood of $(x,y)$. Sub-pixel interface position along the normal is defined so that the interpolated image intensity equals the threshold (Fig. \ref{fig:InterfaceDetection}D-H).

(3) The local tangent angle to the interface $\theta = \tan(\Delta y/\Delta x)$ is found by finite differences, and the contour is re-parameterized with the arc-length parameter $s$.

(4) The total arc length of the interface is measured for each time point. Due to interface deformations, the total arc length fluctuates in time. For computing correlations and Fourier transforms, tangent angle data from a single experiment $\theta(s, t)$ is trimmed so that the interface at each time point has the same total arc length.

\noindent\underline{Power spectra of interface fluctuations.} 

Interface tangent angles $\theta(s,t)$ were used to compute the spatiotemporal autocorrelation function $R_\theta (\Delta s,\Delta t) = \langle \theta(s+\Delta s,t+\Delta t)\theta(s,t)\rangle$, where $\langle\rangle$ denotes averaging over an arc-length interval of 3.1 mm and time interval of 2 hours. The power spectrum was then computed as $S(k) = \int ds e^{-iks} R_\theta (s,\Delta t =0)$. 

\noindent\underline{Dynamic structure factor (DSF) of interface height.}

Local interface height was sampled as a function of the horizontal coordinate $x$ and time $t$. The spatiotemporal autocorrelation function of interface height $R_h (\Delta x,\Delta t) = \langle h(x+\Delta x,t+\Delta t)h(x,t)\rangle$ was calculated by cross correlating a rectangular window $\Delta x_m<x<X-\Delta x_m,\Delta t_m<t<T-\Delta t_m$ of $h(x,t)$ with the complete sample, where $X$ and $T$ denote the sample size and duration. $\Delta x_m$ and $\Delta t_m$ denote maximum lag distance and time respectively. The values of the parameters were $X = 10.7 ~mm, T=2 ~hr,\Delta x_m = 1mm , \Delta t_m  = 0.36 hr$ for experiments; for simulations they were $X = 2 ~mm, \Delta x_m = 0.67~mm$, and $T = 0.44,0.22,0.11 ~hr, \Delta t_m = 0.3,0.15,0.0755 ~hr$ for activity values $\alpha = 10,20,40 \,\mathrm{mPa}$ respectively. The DSF was then computed by multiplying the result of auto-correlation with a 2D Hanning window, and taking the Fourier transform of the result in both space and time.

\noindent\underline{Extraction of the wave dispersion relation.}

Wave dispersion $\omega_p(k)$ was extracted from the DSF in two ways. Above a wave-number $k_{min}$, the DSF for a constant $k$ exhibited a peak at a frequency $\omega = \omega_p(k)$. The peak position was detected by modelling the DSF with a sum of Lorentzians, one centered at $\omega =0 $ and the other at $\omega = \omega_0$:\newline $F(\omega) = a\left((1-c)/((\omega /b)^2+1)+c/((\omega/\omega_0)^2-1)^2+(\omega \Delta \omega/\omega_0^2)^2\right)$, where $a,b,c,\omega_0$ and $\Delta\omega$ are adjustable parameters. The frequency peak was  $\omega_p = \sqrt{\omega_0^2-\Delta\omega^2/2}$. For $k<k_{min}$, the dispersion relation was detected by finding the wave number at which the DSF is at maximum for a constant frequency. 

\noindent\underline{Power spectra of active fluid velocity.} 

To find the velocity of the active phase, two-color fluorescent images were taken of both the Alexa Fluor 488 labeled dextran and the Alexa Fluor 647 labeled microtubules. A mask of the active phase was found by thresholding the dextran channel. This mask, along with the accompanying microtubule images, were imported into the MATLAB plugin PIVLab~\cite{Thielicke2021}. Particle Image Velocimetry was preformed on the microtubule bundle images to find the velocity of the active phase.

Active bulk fluid velocity was calculated in a 2.5 mm X 2 mm window that was 100 microns below the interface. The vectorial velocity field $\mathbf{v}$, sampled in 10 sec intervals over 1.5 hours,  5 hours after the beginning of the experiment, was used to obtain the spatiotemporal autocorrelation function $R_{\mathbf{v}} (\Delta \mathbf{r}, \Delta t) = \langle \mathbf{v}(\mathbf{r}+\Delta\mathbf{r},t+\Delta t)\cdot \mathbf{v}(\mathbf{r},t)\rangle_{\mathbf{r},t}$, where the radius is $\mathbf{r} = (x,y)$. The autocorrelation was azimuthaly averaged to produce the radial autocorrelation function $R_\mathbf{v}(r,t)$. Fig. S13 depicts sections of $R_\mathbf{v}(r,t)$ at $t=0$ and $r=0$ to extract correlation length and time scales.

\noindent\underline{Measuring the center-of-mass of active fluid capillary rise.} 

In wetting experiments and simulations, the center-of-mass height of the active fluid that is adjacent to the wall is defined as follows: 
(1) Interface profile is detected using thresholding as in Fig. \ref{fig:InterfaceDetection}. The average height $Y_0$ of the bulk interface far from the wall ($>5 l_e$) is set as zero height. 
(2) Pixels above threshold whose height is greater than $Y_{0}$, and are within $5 l_e$ of the wall are included in the center-of-mass height determination. The center of mass is defined as $Y_{cm} = \frac{1}{N}\sum_i (Y_i - Y_0)$, where N is the total number of pixels.\\

% ======================================================================

\noindent\textbf{Numerical \red{model}}\\

%\subsection{Multiphase hydrodynamic model \label{sec:SimuMode}}
\noindent\underline{Multiphase hydrodynamic model.}
To simulate  activity-powered interfaces, we use a VOF (Volume Of Fluid)  multiphase hydrodynamic theory to model the active-passive mixture \cite{Hirt1981,Brackbill1992}. The two fluids are described by three continuum fields: the volume fraction of the active phase $\phi$ which is referred to as the `color function' in VOF, the velocity field $\textbf{v}$, and the nematic tensor $\textbf{Q}\equiv S(\hat{\textbf{n}}\hat{\textbf{n}}-\textbf{I})$ describing the local orientation of microtubule bundles. Here, $\hat{\textbf{n}}$ is a unit vector indicating the local orientation of microtubules, and $0\leq S\leq 1$ is the local nematic order parameter. $\textbf{I}$ is the identity matrix. Since the experimental system is quasi-two-dimensional, we \mcm{implement the theoretical model in two dimensions}. The governing  equations are \cite{Hirt1981,Brackbill1992,Marchetti2013,Giomi2014,Blow2014}:
\begin{subequations}
\label{eq:ContEqus}
\begin{align}
 \label{eq:dtphi}
 \frac{D\phi}{Dt}=&0, \\
 \label{eq:dtQ}
\frac{D\textbf{Q}}{Dt}=&\lambda \phi\textbf{u} + \textbf{Q}\cdot\boldsymbol{\omega}-\boldsymbol{\omega}\cdot\textbf{Q} + \red{\frac{1}{\gamma_Q}}\textbf{H}, \\
 \label{eq:dtv}
 \frac{D\rho\textbf{v}}{Dt}=& \eta\nabla^{2}\textbf{v}-\boldsymbol{\nabla} P + \boldsymbol{\nabla} \cdot (\phi\boldsymbol{\sigma})-\gamma_v\mathbf{v}+\textbf{f}_{c}+\textbf{f}_{g},
\end{align}	
\end{subequations}
with $D/Dt=\partial_{t}+\textbf{v}\cdot \boldsymbol{\nabla}$ the material derivative. 

The field $\phi$ is set by the initial conditions to have constant value in the bulk of either phase, with $\phi=1$ in the active fluid and $\phi=0$ in the passive one, and sharp yet continuous variations between the two bulk values at the interface. The advection of $\phi$ by the flow then drives interface fluctuations. \mcm{Unlike the more familiar phase field model, the VOF model sets the right-hand-side of Eq.~\eqref{eq:dtphi} equal to zero, hence neglects \red{phase field} diffusion in the interfacial region. It is appropriate when phenomena such as Ostwald ripening are much slower than other time scales, as appears to be the case in the active-LLPS. It is much more efficient for simulating large interfaces as it only requires interfacial widths of the order of 2-3 grid points.} 

The dynamics of the nematic tensor $\textbf{Q}$ is governed by relaxation and coupling to flow. The first term on the right hand side of Eq.~\eqref{eq:dtQ} describes alignment with local flow gradients, with $\textbf{u}=(\boldsymbol{\nabla}\textbf{v}+\boldsymbol{\nabla}\textbf{v}^{T})/2$ and $\lambda$ the flow-alignment parameter. The flow alignment term is known to drive nematic order even when the system is in the isotropic state \cite{Srivastava2016,Santhosh2020}. To \mcm{confine} this effect \mcm{to the active phase}, we weight \red{the flow-alignment term} by the color function $\phi$ such that flow alignment vanishes \red{in the passive phase}. The second and third term describe co-rotation of the director with the local vorticity $\boldsymbol{\omega}=(\boldsymbol{\nabla}\textbf{v}-\boldsymbol{\nabla}\textbf{v}^{T})/2$, and we neglect for simplicity other nonlinear flow couplings. The relaxation of $\textbf{Q}$, with $\red{\gamma_Q}$ the rotational viscosity, is driven by the molecular field $\textbf{H}=-\delta F_{LdG}/\delta \textbf{Q}$ that minimizes the Landau-de Gennes free energy \cite{DeGennes1993,Marchetti2013}
\begin{equation}
\label{eq:FLdG}
F_{LdG}=\frac{1}{2}\int_{\textbf{r}} \left[a\tr \textbf{Q}^{2}+\frac{1}{2}b\left( \tr \textbf{Q}^{2} \right)^{2}+ K(\partial_jQ_{ik})^{2} \right].
\end{equation}
The first two terms in $F_{LdG}$ control the isotropic-nematic transition, \red{and sets the equilibrium value of order parameter to be} $S=0$ when $a>0$ and $S=\sqrt{-2a/b}$ if $a<0$. The last term describes the energy cost for spatial variation of the order parameter, with isotropic stiffness $K$. \mcm{Here we choose $a>0$. This places the  liquid crystal in the isotropic state when passive, which is the experimentally relevant situation.} 

The velocity field is governed by the Navier-Stokes equation Eq. \eqref{eq:dtv}, with  viscous dissipation controlled by viscosity $\eta$, drag \mcm{$\red{\gamma_v}$ with the walls}, gravitational force $\textbf{f}_{g}=-\rho(\phi) g\hat{\textbf{y}}$, and a capillary force, $\textbf{f}_{c}=\gamma\kappa\boldsymbol{\nabla} \phi$ \cite{Hirt1981,Brackbill1992}, where $\gamma$ is the interfacial tension and $\kappa=-\boldsymbol{\nabla}\cdot (\boldsymbol{\nabla} \phi/|\boldsymbol{\nabla} \phi|)$ the local curvature of the interface. Integrating such capillary force along the interface normal $\hat{\textbf{N}}=\boldsymbol{\nabla} \phi/|\boldsymbol{\nabla} \phi|$ gives a total force $\gamma \kappa \hat{\textbf{N}}$, which is what we expected from Young-Laplace pressure. The pressure $P$ serves as a Lagrange multiplier to incorporate the incompressibility constraint, $\boldsymbol{\nabla}\cdot \textbf{v}=0$. The additional stress from the nematic degrees of freedom,
$\boldsymbol{\sigma}=\boldsymbol{\sigma}^{e}+\boldsymbol{\sigma}^{a}$ includes passive elastic and active stresses, with
\begin{equation}
\label{eq:sigmae}
\boldsymbol{\sigma}^{e}=-\lambda \textbf{H}+\textbf{Q}\cdot\textbf{H}-\textbf{H}\cdot\textbf{Q}\;,~~~~\boldsymbol{\sigma}^{a}=\alpha\textbf{Q}\;,
\end{equation}
and $\alpha\red{<0}$ the activity. Note that $\boldsymbol{\sigma}$ is weighted by the color function $\phi$ in Eq. \eqref{eq:dtv}, hence its contribution vanishes in the passive phase. Similarly, the capillary force $\textbf{f}_{c}$ is nonzero only at the interface. 
For simplicity, we assume that the two phases have the same viscosity and  drag. Finally, the local density is related to the volume fraction $\phi$ as $\rho=\phi\rho_{a}+(1-\phi)\rho_{p}$, where $\rho_{a}$ and $\rho_{p}$ are the densities of pure active and passive phase, respectively. \\

\noindent\textbf{Numerical simulations}\\

%\subsection{Numerical simulations \label{sec:NumeSimu}}
\noindent\underline{General setting.} 

The continuum equations are solved with the Finite Volume \red{M}ethod using the open source package OpenFOAM \cite{Moukalled2016} (OpenFoam, \textit{https://openfoam.org/}). Specifically, we modify the InterFoam solver from OpenFOAM to include the dynamics of the nematic tensor $\textbf{Q}$ \cite{Deshpande2012} (InterFoam, \textit{https://openfoamwiki.net/index.php/InterFoam}). The simulation is done on a square grid embedded in a rectangular box centered at the origin, and we use the standard adaptive time step controller in OpenFOAM with a maximum Courant number $0.3$. Although OpenFOAM can only process three-dimensional simulations, one can still use it to simulate two-dimensional systems by having a single grid along the third dimension, and setting the two boundaries normal to the third dimension to be \textit{empty} (Openfoam user guide, \textit{https://cfd.direct/openfoam/user-guide/}).

The parameters used in the simulations are: $\rho_{a}=1027kg/m^{3}$, $\rho_{p}=1014kg/m^{3}$, $\eta=0.015 Pa\cdot S$, $\gamma_{v}=25 M Pa\cdot S/m^{2}$, $\gamma=0.3 \mu N/m$, $\gamma_{Q}=0.1 kg/(m\cdot S)$, $K=5\times 10^{-14} N$, $a=0.001 Pa$, $b=0.1 Pa$, $\lambda=0.1$. Activity values range from \red{$5$ mPa to $80$ mPa}. The boundary and initial conditions, box and grid sizes \mcm{are varied depending} on the specific problem we study. \\

\noindent\underline{Simulations \mcm{of} interfacial fluctuations.}

We use a rectangular box of size $2mm\times 1mm$ in the $xy$ plane, with a uniform grid spacing of $2.5 \mu m$. \mcm{In the absence of active fluctuations, the interface  separating the top passive fluid from the bottom active fluid is flat and located at $y=0$.}  The top and bottom boundaries are solid walls with slip boundaries for the velocity field $\textbf{v}$, \red{i.e., \MCM{$\hat{\textbf{n}}\cdot\mathbf{v}=0$} and $\partial_{\hat{\textbf{t}}}v_{\hat{\textbf{t}}}=0$ where $\hat{\textbf{n}}$ and $\hat{\textbf{t}}$ represent the normal and tangential directions to the wall,} and \mcm{Neumann}  boundary for the color function $\phi$ and the nematic tensor $\textbf{Q}$, \mcm{i.e., $\nabla\phi|_{wall}=0$ and $\nabla Q_{ij}|_{wall}=0$.} Although no-slip condition is typically used at solid walls, our experiments have found obvious sliding of microtubules with respect to the wall, hence justifying the slip boundary condition of velocity at the wall. At the left and right boundaries \mcm{we impose periodic boundary conditions} by using the \textit{cyclicAMI} boundary in OpenFoam. All simulations start with a flat interface located at $65\%$ of the box height, with zero velocity and zero nematic order. We add small perturbations to the initial $\textbf{Q}$ field in the active phase. Activity then drives these initial perturbations to grow and pushes the system into the chaotic state. \\

\noindent\underline{Simulations \mcm{of} wetting.} 

For the wetting simulations, we use a smaller simulation box of size \red{$0.5 mm\times 0.5 mm$} since we need to use finer grids here. \mcm{The boundary conditions at the top and bottom boundaries are the same as used} in the simulations of interfacial fluctuations. The left and right boundaries are treated as solid walls, with slip boundary conditions for the velocity field and zero-gradient boundary conditions for the color function \mcm{($\nabla\phi|_{wall}=0$)}, except at the \mcm{interface} contact point, where the gradient of $\phi$ \mcm{is set to prescribe the contact angle of the passive system} using the \textit{constantAlphaContactAngle} function in OpenFoam. The nematic tensor $\textbf{Q}$ has a fixed value at the left and right walls. For parallel anchoring of nematic director, we set $Q_{xx}=-0.5$ and $Q_{xy}=0$ at the two walls, and $Q_{xx}=0.5$ and $Q_{xy}=0$ for perpendicular wall anchoring. We use nonuniform grids in the wetting simulations. The grid size in the bulk ($|x|<0.23 mm$) is set to be $2.5 \mu m$ as in the fluctuation simulations. To improve the spatial resolution at the contact point, we refine the simulation grid close to the wall such that the grid size is $1.25\mu m$ for grids within \red{$0.23mm<|x|<0.24mm$} and $0.625\mu m$ for grids \red{at $|x|>0.24mm$}. The initial condition is similar to that in the fluctuation simulations, except the flat interface is located at $50\%$ of the box height. \\

\noindent\underline{Extracting interface profiles.}

The \red{instantaneous} interface profile \red{$h(x,t)$} is extracted from the spatial distribution of color function $\phi$ by using the \textit{isoSurface} function in OpenFOAM. Specifically, OpenFOAM first interpolates among discrete $\phi$ values residing on grids to get a continuously varying $\phi$ field. Based on this, it is able to find numerically the positions where the continuous $\phi$ field is exactly $0.5$. The coordinates of the points with $\phi=0.5$ then constitute the interface profiles we are looking for.\\

\noindent\textbf{\red{Theory of active interfacial fluctuations}.}\\

As discussed in the main text, the non-monotonic power spectra of interfacial tangent angle fluctuations is the result of a competition between passive relaxations and active excitations (Fig. 2B, \red{3B}). Here, we provide the theoretical basis for this claim by analytically deriving \red{the height equation for fluctuating interfaces from continuum hydrodynamics and, based on that, calculate} the fluctuation spectrum of passive and active interface. 
%The geometry is as described in Fig. \ref{fig:tauHSketSI}A. 
We begin by recalling the behavior of thermally driven interfaces.\\

\noindent\underline{Equilibrium interfacial fluctuations from the equipartition theorem.} 

For a system in thermal equilibrium the equal-time spectrum of fluctuations is easily obtained from the free energy cost of  distortions of the flat interface located at $y=0$. We expand the distortion $h(x,t)$ in a Fourier series as $h(x,t)=\frac{1}{L}\sum_k \hat{h}(k,t)~e^{ikx}$, with $L$ the system size along $x$ and inverse transform
$\hat{h}(k,t)=\int_{\mcm{-L/2}}^{\mcm{L/2}}dx\ h(x,t)~e^{-ikx}$.
Assuming small deformations, the free energy cost of interface fluctuations is 
\begin{equation}
    \mathcal{F}=\frac{1}{2L}\sum_k(\gamma k^2+\Delta\rho g)|\hat{h}(k,t)|^2\;,
\end{equation}
where we have included the energy cost due to gravity. Here $\gamma$ is the interfacial tension and $\Delta\rho$ is the difference between the densities of the bottom and top fluid. The equipartition theorem states that the mean energy of each fluctuation mode is $k_BT/2$. It immediately follows
\begin{equation}
\label{eq:hk2Equi}
\frac{1}{L}\langle|\hat{h}(k,t)|^{2}\rangle=\frac{k_{B}T}{\gamma(k^{2}+\ell_{c}^{-2})}\;.
\end{equation}
As is well known, the spectrum becomes constant in the gravity-dominated region $k\ll\ell_c^{-1}$, and scales as $k^{-2}$ at $k\gg\ell_c^{-1}$ where interfacial tension dominates \cite{Safran2003}. \\

\noindent\underline{Equilibrium interfacial fluctuations from interface dynamics.}

Active systems cannot be described by a free energy and require an approach based on dynamics. To set the stage for the study of active interfacial fluctuations, it is useful to first derive  \mcm{ the thermal fluctuation spectrum from hydrodynamics for the case where fluid dissipation is controlled by \emph{both} friction with a substrate and fluid viscosity, as relevant to our experimental system. This derivation, which is not available in the literature, will inform the calculation of the active fluctuation spectrum.}

\mcm{We consider two passive fluids.}
%as in the geometry shown in Fig. \ref{fig:tauHSketSI}A}. 
\mcm{For simplicity assume they have the same viscosity and friction, and only differ in density. We consider the dynamics in the  Stokes limit which is appropriate for our experiments and follow the derivation of Refs. \cite{Grant1983,Harden1991,Flekkoy1995,Flekkoy1996,Bray2001}. The Stokes equation for the two semi-infinite fluids in the presence of thermal noise is given by}
%equation subject to stochastic forcing. This has been done for viscous fluids, and we will now also include substrate friction. Consider an infinite two-dimensional system, where a nearly flat liquid-liquid interface is located at $y=0$. The bulk fluids are governed by the Stokes equation:
%\begin{equation}
%\label{eq:StokesE}
%\eta\nabla^2\textbf{v}-\nabla P-\gamma_v \textbf{v}-\rho g\hat{\textbf{y}}+\nabla\cdot\boldsymbol{\sigma}+\textbf{f}=0,
%\end{equation}
\begin{equation}
\label{eq:StokesE}
\mcm{\gamma_v \textbf{v}=\eta\nabla^2\textbf{v}-\bm\nabla P-\rho g\hat{\textbf{y}}+\mathbf{f}(\mathbf{r},t)\;,}
\end{equation}
\mcm{where} 
\begin{equation}
\mcm{ \mathbf{f}(\mathbf{r},t)= \bm\nabla\cdot\boldsymbol{\sigma}(\mathbf{r},t)+\bm\eta(\mathbf{r},t)}
\end{equation}
\mcm{comprises the  stochastic stress and force density describing the effect of thermal noise, with correlations determined by the fluctuation-dissipation theorem as}
\begin{equation}
\label{eq:CorrFuncSF}
\begin{split}
    \langle \sigma_{ik}(\mathbf{r},t)\sigma_{jl}(\mathbf{r}',t')\rangle=&2k_BT\eta(\delta_{ij}\delta_{kl}+\delta_{il}\delta_{jk})\delta(\mathbf{r}-\mathbf{r}')\delta(t-t')\;, \\
    \langle \eta_{i}(\mathbf{r},t)\eta_{j}(\mathbf{r}',t')\rangle=&2k_BT\gamma_v\delta_{ij}\delta(\mathbf{r}-\mathbf{r}')\delta(t-t')\;,\\
\langle \sigma_{ik}(\mathbf{r},t)\eta_{j}(\mathbf{r}',t'\rangle=&0\;.
\end{split}
\end{equation}
We assume the fluids to be incompressible, hence  $\bm\nabla\cdot\textbf{v}=0$. 
%Here, $\boldsymbol{\sigma}$ and $\textbf{f}$ are stochastic stresses and forces originating from thermal fluctuation. Fluctuation-dissipation theorem demands that they have the following correlations:
%\begin{equation}
%\label{eq:CorrFuncSF}
%\begin{split}
 %   \langle \sigma_{ik}(x,y,t)\sigma_{ik}(x',y',t')\rangle=&2k_BT\eta\delta(x-x')\delta(y-y')\delta(t-t')(\delta_{ij}\delta_{kl}+\delta_{il}\delta_{jk}) \\
%    \langle f_{i}(x,y,t)f_{j}(x,y,t')\rangle=&2k_BT\gamma_v\delta(x-x')\delta(y-y')\delta(t-t')\delta_{ij},
%\end{split}
%\end{equation}
%where $\delta(x)$ is the Dirac delta function and $\delta_{ij}=1$ for $i=j$, and $\delta_{ij}=0$ otherwise. 
Continuity of velocity and stress at the interface requires
\begin{equation}
\label{eq:BounCond}
\begin{split}
    &[\textbf{v}]_0=0\;, \\
    &\left[\eta(\partial_xv_y+\partial_yv_x)+\sigma_{xy} \right]_0=0\;, \\
    &\left[2\eta\partial_yv_y-P+\sigma_{yy} \right]_0=\gamma\partial_x^2h-\Delta\rho gh\;,
\end{split}
\end{equation}
where for any function $s(x,y)$ we have defined $[s(x)]_0\equiv s(x,y=0^-)-s(x,y=0^+)$ as the change in $s$ across the interface. The two terms on the RHS of the last equation represent the Laplace pressure due to interfacial tension and gravity-induced pressure difference, respectively. 

To linear order, the interface height is related to the local velocity through
\begin{equation}
    \partial_th(x,t)=v_y(x,y=0,t)\;.
\end{equation}
To obtain an equation for the dynamics of  interface fluctuations, we need to solve for $v_y(x,y,t)$ for given stochastic force $\mathbf{f}(\mathbf{r},t)$. Interface height correlations will then be obtained by  averaging over thermal noise.

\mcm{By taking Fourier transforms of Eqs. \eqref{eq:StokesE}-\eqref{eq:BounCond} with respect to $x$} and eliminating $\hat{v}_x$ and $\hat{P}$ in favor of $\hat{v}_y$, \mcm{we obtain an equation for}  $\hat{v}_y(k,y,t)$ as
\begin{equation}
\label{eq:PDEvy}
    (\partial_y^2-k^2)(\partial_y^2-\beta^2k^2)\hat{v}_y(k,y,t)=
   \mcm{ \frac{ik}{\eta}\left(\partial_y\hat{f}_x-ik\hat{f}_y\right)}\;,
    %\eta^{-1}k^2\partial_y(\hat{\sigma}_{yy}-\hat{\sigma}_{xx})+\eta^{-1}ik(\partial_y^2+k^2)\hat{\sigma}_{xy}+\eta^{-1}ik\partial_y\hat{f}_x+\eta^{-1}k^2\hat{f}_y,
\end{equation}
where $\beta=\sqrt{1+1/(\ell_{\eta}^{2}k^{2})}$, and $\ell_{\eta}=\sqrt{\eta/\gamma_v}$ is the viscous screening length.
\mcm{We similarly eliminate $\hat{v}_x$ and $\hat{P}$ from Eqs.~\eqref{eq:BounCond} to express the boundary conditions in terms of $\hat{v}_y$, with the result} 
\begin{equation}
\label{eq:BCvy}
\begin{split}
    &[\hat{v}_y]_0=0\;, \\
    &[\partial_y\hat{v}_y]_0=0\;, \\
    &[\eta(\partial_y^2+k^2)\hat{v}_y-ik\hat{\sigma}_{xy}]_0=0\;, \\
    &\left[\frac{\eta}{ k^{2}}(\partial_y^2-3k^2-\ell_{\eta}^{-2})\partial_y\hat{v}_y-\hat{\sigma}_{yy}\mcm{-\frac{i}{k}\hat{f}_x} \right]_0=\gamma(k^2+\ell_c^{-2})\hat{h}\;.
\end{split}
\end{equation}
 \mcm{We write the solution to Eq. \eqref{eq:PDEvy} with boundary conditions given by Eq.~\eqref{eq:BCvy}  as the sum of the solution to the homogeneous equation with the required boundary conditions and a particular solution to the inhomogeneous equation with homogeneous boundary conditions, }
 \begin{equation}
     \hat{v}_y(k,y,t)=\hat{v}_y^h(k,y,t)+\hat{v}'_y(k,y,t)\;.
 \end{equation}
\mcm{The homogeneous solution $\hat{v}_y^h$ describes the flow induced by the stress discontinuity  across the interface and propagated by passive processes. It is given by}
\begin{equation}
\label{eq:vyh}
\begin{split}
\hat{v}_y^h(k,y,t)=&-\frac{|k|}{2\gamma_v}\left[\gamma(k^2+\ell_c^{-2})\hat{h}+[\hat{\sigma}_{yy}+\frac{i}{k}\hat{f}_x]_0\right]\left(e^{-|ky|}-\frac{1}{\beta}e^{-\beta|ky|}\right)\notag\\
&+\frac{ik\hat{\sigma}_{xy}}{2\gamma_v}\textrm{sign}(y)\left(e^{-|ky|}-e^{-\beta|ky|}\right).
\end{split}
\end{equation}
The particular solution can be obtained in terms of Green's function and  describes interfacial flows driven by stochastic stress and force in the bulk. It is given by
\begin{equation}
\label{eq:vya}
\begin{split}
\hat{v}_y'(k,y,t)=&\frac{|k|}{2\gamma_v}\int_{-\infty}^{0_-}dy'\left(e^{-|k||y-y'|}-\frac{1}{\beta}e^{-\beta|k||y-y'|}\right)\left[\hat{f}_y(k,y',t)+\frac{i}{k}\partial_{y'} \hat{f}_x(k,y',t)\right] \\
&+\frac{|k|}{2\gamma_v}\int_{0^+}^{\infty}dy'\left(e^{-|k||y-y'|}-\frac{1}{\beta}e^{-\beta|k||y-y'|}\right)\left[\hat{f}_y(k,y',t)+\frac{i}{k}\partial_{y'} \hat{f}_x(k,y',t)\right].
\end{split}
\end{equation}
%\begin{equation}
%\label{eq:vya}
%\begin{split}
%\hat{v}_y'(k,y,t)=&\frac{|k|}{2\gamma_v}\int_{-\infty}^0dy'\left(e^{|k||y-y'|}-\beta^{-1}e^{\beta|k||y-y'|}\right)\left[\partial_y(\hat{\sigma}_{yy}(y')-\hat{\sigma}_{xx}(y'))+ik^{-1}(\partial_y^2+k^2)\hat{\sigma}_{xy}(y')+ik^{-1}\partial_y\eta_x(y')+\eta_y(y')\right] \\
%&+\frac{|k|}{2\gamma_v}\int_{0}^{\infty}dy'\left(e^{-|k||y-y'|}-\beta^{-1}e^{-\beta|k||y-y'|}\right)\left[\partial_y(\hat{\sigma}_{yy}(y')-\hat{\sigma}_{xx}(y'))+ik^{-1}(\partial_y^2+k^2)\hat{\sigma}_{xy}(y')+ik^{-1}\partial_y\eta_x(y')+\eta_y(y')\right].
%\end{split}
%\end{equation}
Adding the two solutions \red{$\hat{v}_y^h$ and $\hat{v}_y'$},  and integrating by parts, it is easy to show that, at the interface $y=0$, the surface terms in $\hat{v}'_y$ are cancelled by equal and opposite contributions from $\hat{v}^h_y$. The $y$ velocity at the interface can then be written as 
\begin{equation}
    \hat{v}_y(k, y=0,t)=\hat{v}^r(k,t)+\hat{v}^t(k,t)\;,
\end{equation}
where $\hat{v}^r$ controls the passive relaxation of the  interface  due to surface tension and gravity  and $\hat{v}^t$ represents the stochastic forcing arising from thermal noise that drives interface fluctuations. The relaxation part has the form
\begin{equation}
\label{eq:vyr}
\hat{v}^r(k,t)=-\nu(k)\hat{h}(k,t)\;,
\end{equation}
with 
\begin{equation}
\label{eq:nu}
\nu(k)=\frac{\gamma(k^2+\ell_c^{-2})}{2\eta|k|(\beta^2+\beta)}\equiv\frac{\gamma(k^2+\ell_c^{-2})}{\zeta(k)}\;,
\end{equation}
where 
\begin{equation}
\zeta(k)=2\eta|k|(\beta^2+\beta)   
\end{equation} 
has a natural interpretation as an effective friction per unit length on the interface.
The stochastic forcing has a rather lengthy expression
\begin{equation}
\label{eq:vyd}
\begin{split}
\hat{v}^t(k,t)=&\frac{k^2}{2\gamma_v}\int_{-\infty}^0dy'e^{|k|y'}\left[\hat{\sigma}_{xx}(y')-\hat{\sigma}_{yy}(y')+2ik|k|^{-1}\hat{\sigma}_{xy}(y')-ik^{-1}\hat{f}_x(y')+|k|^{-1}\hat{f}_y(y')\right] \\
&-\frac{k^2}{2\gamma_v}\int_{-\infty}^0dy'e^{\beta|k|y'}\Big[\hat{\sigma}_{xx}(y')-\hat{\sigma}_{yy}(y')+ik|k|^{-1}(\beta+\beta^{-1})\hat{\sigma}_{xy}(y')-ik^{-1}\hat{f}_x(y')\\
&+\beta^{-1}|k|^{-1}\hat{f}_y(y')\Big] \\
&-\frac{k^2}{2\gamma_v}\int_{0}^{\infty}dy'e^{-|k|y'}\left[\hat{\sigma}_{xx}(y')-\hat{\sigma}_{yy}(y')-2ik|k|^{-1}\hat{\sigma}_{xy}(y')-ik^{-1}\hat{f}_x(y')-|k|^{-1}\hat{f}_y(y')\right] \\
&+\frac{k^2}{2\gamma_v}\int_0^{\infty}dy'e^{-\beta|k|y'}\Big[\hat{\sigma}_{xx}(y')-\hat{\sigma}_{yy}(y')-ik|k|^{-1}(\beta+\beta^{-1})\hat{\sigma}_{xy}(y')-ik^{-1}\hat{f}_x(y')\\
&-\beta^{-1}|k|^{-1}\hat{f}_y(y')\Big]\;,
\end{split}
\end{equation}
%The first(last) two integrals correspond to flow driven by stochastic stress and force above(below) the interface. Here,  [cite Bray and maybe others]
\mcm{but it is easy to show using Eqs.~\eqref{eq:CorrFuncSF}  that  it  has zero mean and   correlations} 
\begin{equation}
\langle \hat{v}^t(k,t)\hat{v}^t(k',t')\rangle=\frac{2k_BT}{\zeta(k)}L\delta_{k,-k'}~\delta(t-t')\;.
\label{eq:noise_t}
\end{equation}

The \mcm{dynamics of interface fluctuations} is then governed by a Langevin equation
\begin{equation}
 \partial_t\hat{h}(k,t)=-\nu(k)\hat{h}(k,t)+\hat{v}^t(k,t)\;. 
 \label{eq:hdot_T}
\end{equation}
with noise correlations given by Eq.~\eqref{eq:noise_t}.
We can now use the Langevin equation  to evaluate the equal time spectrum of interface fluctuations. After defining the temporal Fourier transform, $\hat{h}(k,\omega)=\int_{-\infty}^{\infty} dt e^{-i\omega t}\hat{h}(k,t)$, we immediately obtain the dynamic structure factor of the interface as
\begin{equation}
    \frac{1}{L}\langle |\hat{h}(k,\omega)|^2\rangle=\frac{2k_BT/\zeta(k)}{\omega^2+\nu^2(k)}\;.
\end{equation}
The static or equal-time fluctuation spectrum is then given by
\begin{equation}
\label{eq:H2k}
\frac{1}{L}\langle |\hat{h}(k,t)|^2\rangle=\int_{-\infty}^{\infty}\frac{d\omega}{2\pi L}\langle |\hat{h}(k,\omega)|^2\rangle=\frac{2k_BT/\zeta(k)}{2\nu(k)}=\frac{k_BT}{\gamma (k^2+\ell_c^{-2})}\;,
\end{equation}
which is consistent with that obtained using the equipartition theorem. \mcm{Importantly, the dependence of the noise amplitude on the effective friction is key for guaranteeing the result obtained from equipartition.} \\
%Finally, the two-time correlation function is
%\begin{equation}
%\label{eq:CorrFuncH}
%\langle \hat{h}(k,\tau)\hat{h}(-k,t+\tau)\rangle=\frac{k_BTL}{\gamma (k^2+\ell_c^{-2})}e^{-\nu(k) \tau},
%\end{equation}
%which shows a simple exponential decay with rate $\nu(k)$.\\

\noindent\underline{Dynamics of activity-powered interfacial fluctuations.} 

In the previous section, we studied the dynamics of thermally driven interfaces. This is characterized by the exponential relaxations with $k$ dependent rates that are controlled by interfacial tension and gravity. Now we will show how a non-monotonic tangent angle spectrum arises in active interfaces from the competition of the passive relaxation mechanisms delineated above and active processes. When the bottom fluid is active, the main driving force of interfacial fluctuations is not thermal noise, but active stress. Neglecting random  thermal forces and stresses, the Stokes equation then takes the form
\begin{equation}
\label{eq:StokesA}
\mcm{\gamma_v \textbf{v}=\eta\nabla^2\textbf{v}-\bm\nabla P-\rho g\hat{\textbf{y}}+\bm\nabla\cdot\bm\sigma^a\;,}
\end{equation}
where the active stress $\bm\sigma^a$ is specified below.  %is determined by the familiar active stress,
%\begin{equation}
%  \mathbf{f}^a=\bm\nabla\cdot\bm\sigma^a\;,  
%\end{equation}
%\yzh{
%\MCM{to be specified below.}
%We do not assume any specific dynamics or structure for $\boldsymbol{\sigma}^a$, but only consider it to be a traceless, symmetric tensor. 
%For the active nematic system of interest here, } $\bm\sigma^a=\alpha\mathbf{Q}$, with $\mathbf{Q}$ the nematic alignment tensor and $\alpha$ the activity.

One can then carry out the same derivation as in the thermal case to obtain an  equation for the interfacial fluctuations as
\begin{equation}
 \partial_t\hat{h}(k,t)=-\nu(k)\hat{h}(k,t)+\hat{v}^a(k,t)\;,  
 \label{eq:hdot_A}
\end{equation}
where the passive relaxation rate $\nu(k)$ is given again by Eq. \eqref{eq:nu} and $\hat{v}^a(k,t)$ is the active forcing (or active interfacial flow) due to bulk active stress, given by
\begin{equation}
\label{eq:vakt}
\begin{split}
\hat{v}^a(k,t)=&\frac{k^2}{2\gamma_v}\int_{-\infty}^0dy\ e^{|k|y}\left[2\hat{\sigma}^a_{xx}(y)+2ik|k|^{-1}\hat{\sigma}^a_{xy}(y)\right] \\
&-\frac{k^2}{2\gamma_v}\int_{-\infty}^0dy\ e^{\beta|k|y}\left[2\hat{\sigma}^a_{xx}(y)+ik|k|^{-1}(\beta+\beta^{-1})\hat{\sigma}^a_{xy}(y)\right].
\end{split}
\end{equation}
In this case, however, Eq.~\eqref{eq:hdot_A} is not a closed equation since the forcing $\hat{v}^a$ is determined by the dynamics of \MCM{the active stress,}
%the nematic order parameter $\mathbf{Q}$, 
which in turn couples back to the flow, as shown in Eq.~\eqref{eq:dtQ}. As discussed in the main text, this feedback is key for the onset of traveling surface waves. On the other hand, as shown below, the form given in Eq. ~\eqref{eq:hdot_A}  offers a useful interpretation of the role of activity on the equal-time fluctuation spectrum.

\yzh{
\MCM{To proceed, we treat the active stress as stochastic forcing } 
%Compared to the equilibrium thermal noise, the most distinguished feature of active noise is that it's 
correlated  both in space and time. \MCM{This is justified by a large body of simulations of bulk active nematics~\cite{Doostmohammadi2018} that have quanified active stress correlations. For the purpose of modeling interfacial fluctuations,
%To capture this feature, 
we assume a simple form with exponential \MCM{correlation in both space and time, given by}}
%spatial-temporal correlation for active stress:
%
\begin{equation}
  \label{eq:CFQ}
  \langle \sigma^{a}_{xx}(\textbf{r},t)\sigma^{a}_{xx}(\textbf{r}',t') \rangle=\langle \sigma^{a}_{xy}(\textbf{r},t)\sigma^{a}_{xy}(\textbf{r}',t') \rangle=\sigma_{rms}^2e^{-|\textbf{r}-\textbf{r}'|/\ell_{a}}e^{-|t-t'|/\tau_{a}}\;,
\end{equation}
and $\langle \sigma^{a}_{xx}(\textbf{r},t)\sigma^{a}_{xy}(\textbf{r}',t') \rangle=0$. The statistical properties of the active noise \MCM{are} then completely determined by three quantities: the correlation length $\ell_{a}$, the correlation time $\tau_{a}$, and  root mean square active stress $\sigma_{rms}$. %In active nematics, since $\boldsymbol{\sigma}^{a}=\alpha\textbf{Q}$, we have $\sigma_{rms}^2=\alpha^{2}\bar{S}^{2}/8$, where $\bar{S}$ is the mean nematic order parameter, which can be measured from simulations. 
Using Eqs. \eqref{eq:vakt}-\eqref{eq:CFQ}, we can calculate the correlation function of the active forcing $v^{a}$ as 
\begin{equation}
\label{eq:CFva}
\langle \hat{v}^{a}(k,t)\hat{v}^{a}(k',t')\rangle=\frac{2\mathcal{E}(k)}{\zeta(k)}L\delta_{k,-k'}~\MCM{\frac{e^{-|t-t'|/\tau_{a}}}{\tau_{a}}}\;,
\end{equation}
where
\begin{equation}
\label{eq:EkTheory}
\mathcal{E}(k)=\frac{\sigma_{rms}^2\ell_{a}^{2}\tau_{a}\zeta(k)}{8\gamma_{v}^{2}}\int_{-\infty}^{\infty}dk_{z}\frac{(\beta-1)^{2}k^{6}}{\left( k^{2}+k_{z}^{2} \right)\left(\beta^{2} k^{2}+k_{z}^{2} \right)}\frac{4+\left(1+\beta^{-1}\right)^{2}+\left(1-\beta^{-1}\right)^{2}k_{z}^{2}k^{-2}}{\left( 1+\ell_{a}^{2}k^{2}+\ell_{a}^{2}k_{z}^{2} \right)^{3/2}}.
\end{equation}
We can then readily obtain the equal-time spectrum of the active interfacial fluctuations as
\begin{equation}
\label{eq:PSHa}
\frac{1}{L}\langle |\hat{h}(k,t)|^{2}\rangle=\frac{2\mathcal{E}(k)}{\gamma(k^{2}+\ell_{c}^{-2})}\frac{1}{1+\tau_{a}\nu}.
\end{equation}
%\begin{equation}
%\label{eq:CorrFuncHa}
%\langle \hat{h}(k,\tau)\hat{h}(-k,t+\tau)\rangle=\frac{2\mathcal{E}(k)L}{\zeta(k)}\frac{1}{\nu \left( %1+\tau_{a}\nu \right)}\frac{e^{-\nu(k) |t|}-\tau_a\nu(k)e^{-|t|/\tau_a}}{1-\tau_a\nu(k)},
%\end{equation}
%as well as the equal-time spectrum
%\begin{equation}
%\label{eq:PSHa}
%\frac{1}{L}\langle |\hat{h}(k,t)|^{2}\rangle=\frac{2\mathcal{E}(k)}{\gamma(k^{2}+\ell_{c}^{-2})}\frac{1}{1+\%tau_{a}\nu}.
%\end{equation}
}

%Equation \eqref{eq:CorrFuncHa} suggests that the dynamics of the active interface is controlled by the interplay of passive relaxation at rate $\nu(k)$ and active driving on times scale $\tau_a$. At small $k$, $\nu^{-1}\gg \tau_a$ (Figs. \ref{fig:tauHSketSI}B--C) and passive relaxation dominates. Intuitively, the correlation of active flows are short lived compared to the passive relaxation of long wavelength interfacial fluctuations and resemble thermal noise, albeit with a $k$-dependent spectrum $\sim\mathcal{E}(k)$ that accounts for deviations from equipartition. Conversely, at large $k$, $\tau_a\gg \nu^{-1}$ (Figs. \ref{fig:tauHSketSI}B--C). At short scales the interface relaxes quickly and its dynamics is slaved to the slowly varying active driving. To demonstrate this, we measure the correlation time of active interface from simulations and experiments. Figure \ref{fig:tauHSketSI}B--C clearly show that in both systems $\tau_h(k)$ crossovers from $\nu^{-1}$ to $\tau_a$ as we increase wavenumber $k$.

\yzh{
The equal-time spectrum of the active interface is well described by Eq. \eqref{eq:PSHa}. Figures \ref{fig:pEkHk}A,B show excellent agreement between the theoretical spectra (solid lines) calculated using Eq. \eqref{eq:PSHa} and those measured from simulations and experiment (circles). The experiment, simulations, and theory, all suggest a crossover of the height spectrum from $\langle|\hat{h}(k,t)|^2\rangle\sim|k|$ at small wavenumber to $\langle|\hat{h}(k,t)|^2\rangle\sim k^{-6}$ at large wavenumber, which is very different from the equilibrium spectrum in Eq. \eqref{eq:H2k}. This can be attributed to the scale dependence of energy injection in the active fluid. 

To understand this, note that $\mathcal{E}(k)$ has the dimension of energy. Comparing Eq. \eqref{eq:CFva} to \eqref{eq:noise_t}, we see that $\mathcal{E}(k)$ can be used to characterize the energy scale of active fluctuations, to be compared to $k_{B}T$ in thermal equilibrium. We have calculated $\mathcal{E}(k)$ numerically  using Eq. \eqref{eq:EkTheory} and the results are shown in Figs. \ref{fig:pEkHk}C,D. Both simulations and experiments show energy scale of the order $10^{-13}\sim 10^{-11} J$, which is much larger than the thermal energy scale $k_{B}T\sim 10^{-21}J$. This explains the giant interfacial fluctuations found in both experiments and simulations. Furthermore, $\mathcal{E}(k)$ has a strong dependence on wavenumber $k$: it crosses over from $\mathcal{E}(k)\sim k$ at small $k$ where dissipation is dominated by friction to $\mathcal{E}(k)\sim k^{-3}$ where dissipation is dominated by viscosity. The crossover length scale is essentially independent of activity and is controlled by the typical size of flow vortices, which in our system is determined by the screening length $\ell_\eta$. \MCM{This behavior is consistent with the energy spectrum reported for bulk active liquid crystal in the regime of active turbulence~\cite{Martinez-Prat2021}.}

The \MCM{scale dependence} of the active energy injection determines the fluctuation spectrum of active interface.} At small wavenumber or large scales, 
%where $\tau_a\nu(k)\ll 1$ and passive relaxation controls the interface dynamics 
we find $\langle|\hat{h}(k,t)|^{2}\rangle\sim \mathcal{E}(k)\sim k$, in agreement with the interfacial spectra shown in Figs.~\ref{fig:pEkHk}A,B from both simulations and experiments. At large wavenumber where $\tau_a\nu(k)\gg 1$,
%and active processes control the interface dynamics, 
we find $\langle|\hat{h}(k,t)|^{2}\rangle\sim \mathcal{E}(k)/(k^2\nu(k))\sim k^{-6}$, where we have assumed $k\gg\ell^{-1}_c,\ell_\eta^{-1}$ and used $\nu(k)\sim k$. The scale-dependence of active energy injection distinguishes the active interfacial spectra from their equilibrium counterparts. \\
%For the parameter values used in the experiments, the peak wavenumber in the velocity energy spectrum, denoted by $k_e$ for convenience, is smaller than the $\ell_c^{-1}$.  For $k<k_e$, one can approximate $\mathcal{E}(k)\sim k$ and, subsequently, $L^{-1}\langle|\hat{h}(k,t)|^{2}\rangle\sim k/(k^2+\ell_c^{-2})$, which is peaked at $k_m\approx \ell_c^{-1}$. Figure \ref{fig:pHkSigs}A shows the fluctuation spectra at different interfacial tensions. We extract $k_m$ from these lines, and find that $k_m$ indeed increases linearly with $\ell_c^{-1}$ (Fig. \ref{fig:pHkSigs}B). These results indicate that the crossover length in the interface spectrum is controlled by the capillary length $\ell_c$. \\

\yzh{
\noindent\underline{Estimating active stress using interfacial fluctuations.}

Equation \eqref{eq:PSHa} allows us to estimate the magnitude of the active stress $\sigma_{rms}$ from the interface spectrum.  \MCM{This is best done} using the interface's tilting angle $\theta$ instead of the height since the former is well \MCM{defined} even at high activity. Using $\hat{\theta}(k,t)\simeq ik\hat{h}(k,t)$, we find
\begin{equation}
\label{eq:PSTa}
\frac{1}{L}\langle |\hat{\theta}(k,t)|^{2}\rangle=\frac{2k^{2}\mathcal{E}(k)}{\gamma(k^{2}+\ell_{c}^{-2})}\frac{1}{1+\tau_{a}\nu},
\end{equation}
and the root mean square value of $\theta$
\begin{equation}
\label{eq:RMSTa}
\theta_{rms}\equiv L^{-1}\sqrt{\sum_{k}\langle |\hat{\theta}(k,t)|^{2}\rangle}.
\end{equation}
\MCM{Clearly} $\theta_{rms}$ is proportional to the amplitude of the active stress $\sigma_{rms}$,
\begin{equation}
\label{eq:RMSTa1}
\theta_{rms}=\sigma_{rms}/p,
\end{equation}
where
\begin{equation}
\label{eq:p}
\frac{1}{p^2}=\sum_k\frac{\ell_{a}^{2}\tau_{a}L^{-1}k^{2}}{4\gamma_{v}^{2}\nu(1+\tau_{a}\nu)}\int_{-\infty}^{\infty}dk_{z}\frac{(\beta-1)^{2}k^{6}}{\left( k^{2}+k_{z}^{2} \right)\left(\beta^{2} k^{2}+k_{z}^{2} \right)}\frac{4+\left(1+\beta^{-1}\right)^{2}+\left(1-\beta^{-1}\right)^{2}k_{z}^{2}k^{-2}}{\left( 1+\ell_{a}^{2}k^{2}+\ell_{a}^{2}k_{z}^{2} \right)^{3/2}}.
\end{equation}
Measuring $\theta_{rms}$ and calculating $p$ numerically allows us to estimate the magnitude of active stress $\sigma_{rms}$. 
%To benchmark the accuracy of this method, we estimate $\sigma_{rms}$ and compare it with the result $\sigma_{rms}^0\equiv \alpha^2\bar{S}^2/8$ \MCM{that applies in simulations where $\bm\sigma^a=\alpha\mathbf{Q}$.} Here,  $\bar{S}$ is the mean nematic order parameter which can be measured from simulations. As shown in Fig. xA, this method gives no more than $50\%$ discrepancy from the expected values. 
We use this method to measure the active stress in the experiment. Taking advantage of the fact that $\ell_{a}$ and $\tau_{a}$ barely change with the KSA concentration, $p$ is essentially independent of the KSA concentration. Using $p\approx 8.4$ mPa/rad estimated from one set of data, we obtain that the active stress varies between $2.5$ mPa and $6$ mPa in the experiment (Fig. 5B), close to \MCM{the values obtained} from activity-induced wetting \red{below 300 nM KSA. The lowest active stress value coincides with the yield stress of passive kinesin-crosslinked bundled microtubule gels \cite{Gagnon2020}.}\\ 
}

% ----------------------------------------------------------------------

\noindent\textbf{Theory of activity-induced wetting: from active stress to active tension}\\

The enhanced wetting in the presence of activity originates from directed active stresses in the region \mcm{near} the wall that  persistently lift the interface, effectively increasing wall adhesion of the active phase. \red{Both experiment and s}imulation show that \red{nematic director preferentially} aligns with the wall (Fig. 4B, 4D inset). \mcm{Such an alignment is expected even for passive rigid filaments due to steric interaction with the wall~\cite{DeGennes1993,Kimura1985}. It is enhanced by active forces, resulting in so-called active anchoring, as demonstrated in recent simulations~\cite{Blow2014,Blow2017,Coelho2021}}. Since the active stress is extensile, these vertically aligned domains exert, on average, a lifting force on the interface, driving it upwards. Activity then changes both the height of the contact point and the apparent wetting angle, as shown below. \\

\noindent\underline{\mcm{Force balance at a passive} interface.}

\mcm{We first summarize the force balance that determines the interface profile and}  the wetting \mcm{angle} of a passive interface \mcm{in the presence of}  gravity \cite{Berg2010}. For a passive interface, the  profile \mcm{of the interface height $h(x,t)$} is governed by the Young-Laplace equation \mcm{that expresses normal force balance across the interface as }
\begin{equation}
\label{eq:YLE}
\gamma \frac{h''}{\left( 1+h'^{2} \right)^{3/2}}-\Delta\rho g h=0\;,
\end{equation}
\mcm{where primes denote derivatives with respect to $x$ and  $\gamma$ is the interfacial tension. This equation needs to be solved with the contact}  boundary condition at the wall
\begin{equation}
\label{eq:BCYLE}
h'(0)=-\mcm{\cot}\theta_e\;,
\end{equation}
where \mcm{the wetting angle $\theta_e$ (shown in Fig.~\ref{fig:wet-sket}\red{A})} is determined by balancing the wall tension $\gamma_{w}$ and the interfacial tension $\gamma$
\begin{equation}
\label{eq:theta}
\gamma\cos\theta_e=\gamma_{w}\;.
\end{equation}
\mcm{Assuming the slope of the interface remains small, i.e., $h'\ll 1$, Eq.~\eqref{eq:YLE} can be linearized and solved, resulting in an exponential interface profile, given by}
\begin{equation}
 h(x)=\ell_c\cot\theta_e ~e^{-x/\ell_c}\;,
    \label{eq:int_profile}
\end{equation}
\mcm{with $\ell_c=\sqrt{\gamma/\Delta\rho g}$ the capillary length.}\\

\noindent\underline{\mcm{Force balance at an} active interface.} 

%Now let's see what's happening to an active interface. 
\mcm{In the active liquid crystal there is} a region close to the wall where MT bundles align parallel to the wall (Figs. 4B, 4D inset, ~\ref{fig:wet-sket}B).    We assume that the thickness $\ell_w$ of this \mcm{wall-aligned} region is $\ell_{w}\ll \ell_{c}$. Outside this region ($x>\ell_{w}$), the average active stress vanishes due to the chaotic dynamics, and we expect the average interface profile to be governed again by the Young-Laplace equation,  \mcm{ but with an apparent wetting angle $\theta_{a}$ different from the equilibrium wetting angle $\theta_e$ (Fig.~\ref{fig:wet-sket}B). The profile is therefore given by Eq.~\eqref{eq:int_profile}, with $\theta_e\rightarrow\theta_{a}$.} 
%The Young-Laplace equation, together with the `new' wetting angle $\theta_{a}$, then determines the interface profile except at $x<\ell_{w}$, which is less important as long as $\ell_{w}\ll \ell_{c}$.

To determine $\theta_{a}$, \mcm{we examine force balance within the thin wall-aligned region where four forces per unit chamber thickness are at play (inset Fig.~\ref{fig:wet-sket}B)}:
\begin{enumerate}
\item coherent active stresses lifting the interface  $F_{a}\approx -\alpha\ell_{w}$, where \mcm{$\alpha<0$};
\item vertical downward gravitaional force resulting from density difference,  $F_{g}=\Delta\rho g h_{0}\ell_{w}$, where  $h_{0}$ is the height of the contact point;
\item interfacial tension away from the wall aligned domain at $x>\ell_w$, drags the interface downward  $F_{i}=\gamma \cos\theta_{a}$ \mcm{in the $y$ direction;}
\item wall adhesion contributes to a vertical lifting force per unit length $F_{w}=\gamma_{w}$.
\end{enumerate}
\mcm{The wetting angle $\theta_{a}$ is determined by} the balance of these four terms through
\begin{equation}
\label{eq:EqTheta1}
F_w + F_a =F_i+F_g\;,
\end{equation}
$-\alpha\ell_{w}>0$ has the dimension of interfacial tension which defines an ``active tension'' $\gamma_{a}\equiv |\alpha|\ell_{w}$. 

%Equation \eqref{eq:EqTheta11} describes that with increasing activity, the contact angle $\theta_{a}$ decreases while the maximum height $h_{0}$ increases. 
\mcm{Prior to} complete wetting, \mcm{the interface profile for $x\ge\ell_{w}$ is governed by} the Young-Laplace law \mcm{with wetting angle $\theta_{a}$. \mcm{The interface profile must be obtained from the solution of the nonlinear equation, Eq.~\eqref{eq:YLE}}~\cite{Berg2010}. The maximum height $h_{0}$ is obtained by setting $h(x=0)\simeq h(x=\ell_w)$, with the result} 
\begin{equation}
\label{eq:h0}
h_{0}\approx\ell_{c}\sqrt{2(1-\sin\theta_{a})}.
\end{equation}

Equation \eqref{eq:EqTheta1} then becomes
\begin{equation}
\label{eq:EqTheta12}
\gamma\cos\theta_{a}+\Delta \rho g\ell_{w}\ell_{c}\sqrt{2(1-\sin\mcm{\theta_{a}})}=\gamma_{w}+|\alpha| \ell_w\;.
\end{equation}
This shows that increase in activity results in a decrease of the active wetting angle $\theta_{a}$, and associated increase of the maximum height $h_{0}$.

The onset of complete wetting corresponds to $\theta_{a}=0$. Inserting this in Eq.~\eqref{eq:EqTheta12} gives the critical activity for complete wetting as
\begin{equation}
\label{eq:alphac}
\alpha_{c}=\frac{\gamma-\gamma_{w}}{\ell_w}+\sqrt{2}\Delta\rho g\ell_{c}\;.
\end{equation}
Beyond complete wetting, 
%the average interfacial profile at $x\ge \ell_{w}$ satisfies the Young-Laplace equation with a boundary condition 
$\theta_{a}=0$. For $x\le \ell_{w}$, the interface height keeps growing until the active stress is balanced by gravity and interfacial tension. Setting $\theta_{a}=0$ in Eq. \eqref{eq:EqTheta1}, we obtain the maximum height as
\begin{equation}
\label{eq:h0CompWet}
h_{0}=\frac{1}{\Delta\rho g\ell_w}\left( \gamma_{w}-\gamma-\alpha\ell_w \right)\mcm{=\sqrt{2}\ell_c+\frac{1}{\Delta\rho g}\left(|\alpha|-\alpha_c\right)}\;.
\end{equation}
When $|\alpha|\gg \alpha_{c}$, i.e., \mcm{at values of activity well above} the onset of complete wetting,
\mcm{$h_0\simeq \frac{|\alpha|}{\Delta\rho g}$ and one can infer the active stress directly by measuring $h_0$.} In experiments, we use the center-of-mass of the capillary rise to determine the active stress, as it is challenging to consistently define $h_0$, e.g. when the wetting layer splits from the bulk fluid.

Numerical simulations allowed us to examine the dynamics of active wetting starting from a flat interface. For all activities, the contact point  climbed \mcm{up} the wall, \mcm{saturating} at a maximum value \mcm{determined by  force balance}  (Fig. \ref{fig:h-wet}A). \mcm{We compare the steady state maximum height obtained from simulations (circles) to Eq~\eqref{eq:EqTheta1} (Fig. \ref{fig:h-wet}B).} 
The theory provides an excellent prediction \mcm{for}  the height prior to complete wetting,
as well as the transition to complete wetting. Beyond complete wetting, the maximum height $h_0$  increases linearly with activity $|\alpha|$, suggesting that measuring the height of the contact point estimates active stresss. The rate of growth of the height with activity is, however, slightly larger than the expected value $1/\Delta \rho g$ (dashed line in Fig. \ref{fig:h-wet}B). This could be due to the fact that at high activity, the contact point has a sharp geometry and the fields are varying violently in space, which significantly reduces the numerical accuracy of OpenFOAM.

To summarize, the presence of \mcm{a wall-aligned} layer gives rise to an active tension $\gamma_{a}=|\alpha|\ell_{w}$ that changes the apparent wetting angle from the passive value $\theta_e$ to $\theta_{a}$ given by Eq. \eqref{eq:EqTheta12}. \mcm{Importantly, we show that it is possible to infer the active stress by comparing measurement of  wetting of interfaces in} passive and active samples. \\

% ----------------------------------------------------------------------
%\subsection{Converting maximum height  to center of mass \mcm{of interface} \label{sec:h02hc}}

\noindent\underline{Converting maximum height  to center of mass.}

\mcm{In experiments it is difficult to determine the maximum height of the wetting layer. A more convenient and directly measurable quantity for  quantifying  activity-induced wetting is} the center of mass of the region \mcm{of the interface that is lifted above its flat value, defined as} 
\begin{equation}
\label{eq:hc}
h_{c}=\frac{1}{2A}\int_{0}^{\infty} h^{2}(x)dx\;,
\end{equation}
where
\begin{equation}
\label{eq:A}
A=\int_{0}^{\infty} h(x)dx
\end{equation}
is the area of the lifted region. \mcm{To compare with experiments we therefore evaluate the center of mass $h_c$ as follows. By requiring that}  the total gravitational force exerted on the lifted region, $\Delta\rho g Ad$,  balance the \mcm{sum of} wall adhesion $F_{w}$ and active lifting force $F_{a}$, we  obtain the area $A$ as
\begin{equation}
\label{eq:A1}
A=\frac{\gamma_{w}-\alpha\ell_{w}}{\Delta\rho g}\;.
\end{equation}
\mcm{To evaluate t}he integral in Eq. \eqref{eq:hc}. \mcm{we  separate} the wall-aligned layer from the bulk part of the interface. Within the aligning layer $x<\ell_{w}$, we assume  \mcm{$h(x)\simeq h_{0}$, so that} the thin layer  contributes $h_{0}^{2}\ell_{w}/2$ to the integral. For $x\ge \ell_{w}$, we assume the average interface profile  satisfies the \mcm{nonlinear} Young-Laplace equation. \mcm{Since} no explicit \mcm{solution is available,}  we first evaluate the interface profile numerically by solving the Young-Laplace equation with contact angle $\theta_{a}$, then calculate the bulk part of the integral in Eq. \eqref{eq:hc} numerically. \\

\clearpage

\begin{figure}[t]
\centering
\includegraphics[width=\textwidth]{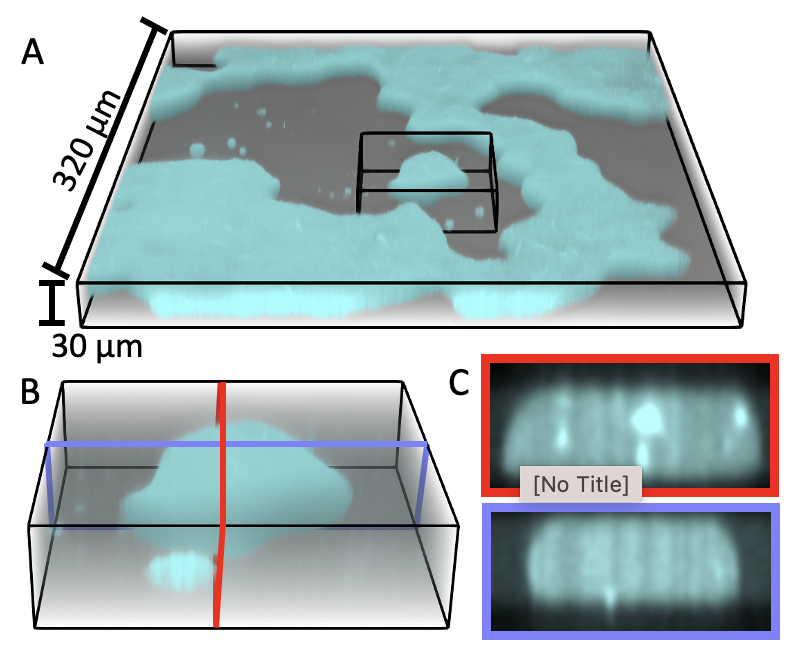}
\caption{\label{fig:droplets} \textbf{Three-dimensional visualization of coarsening sample at 275 nM KSA.} (\textbf{A}) Active droplets (cyan) confined to a 30 µm chamber. (\textbf{B}) Magnified image of an isolated droplet. (\textbf{C}) Cross sections views of the droplet. In 30 µm thick chambers, droplets span the entire chamber, and have a nearly flat vertical profile.}
\end{figure}

\clearpage

\begin{figure}[t]
\centering
\includegraphics[width=\textwidth]{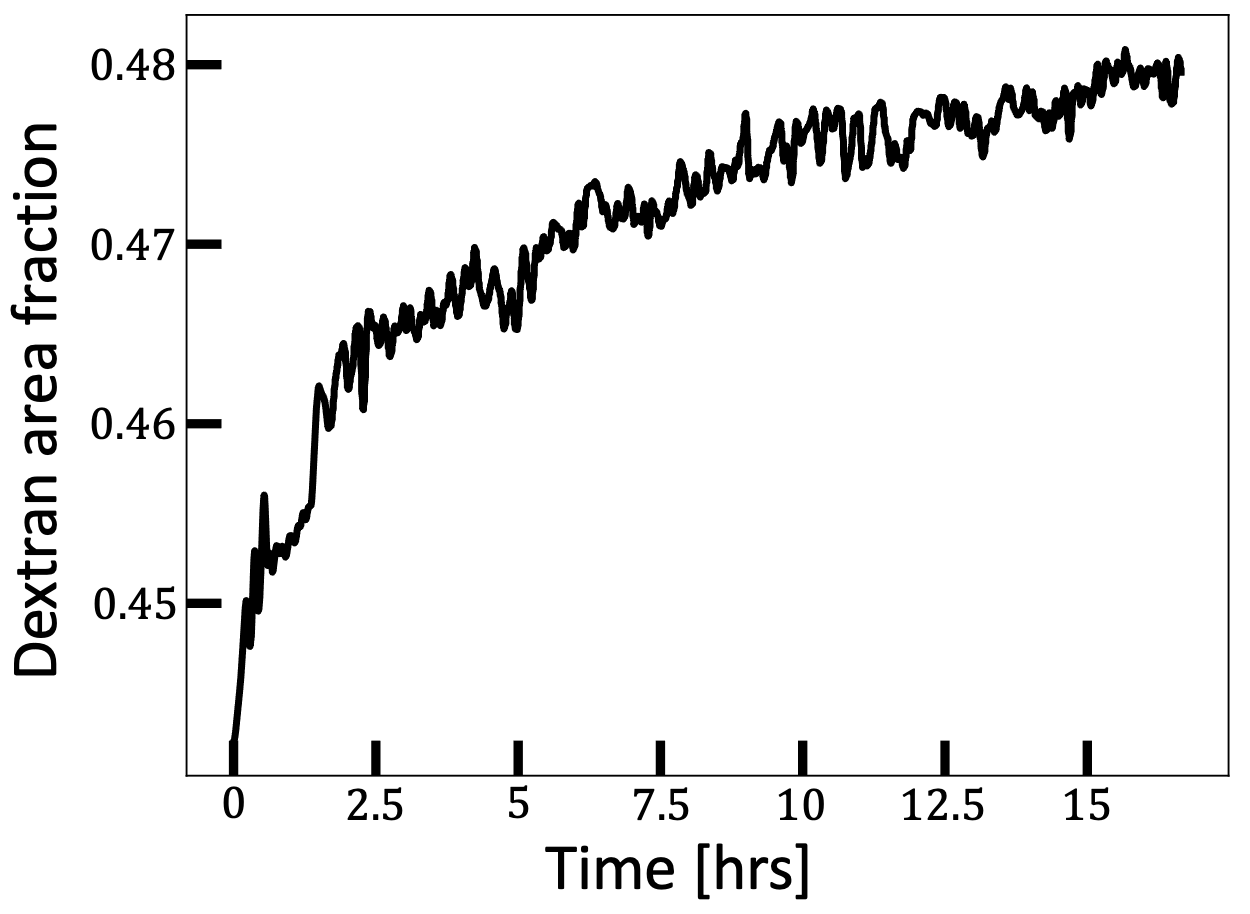}
\caption{\label{fig:AreaFraction} \textbf{Area fraction of dextran over time.} The area fraction initially increases rapidly, then remains nearly constant 2 hours after sample preparation.  
}
\end{figure}

\clearpage

\begin{figure}[t]
\centering
\includegraphics[width=\textwidth]{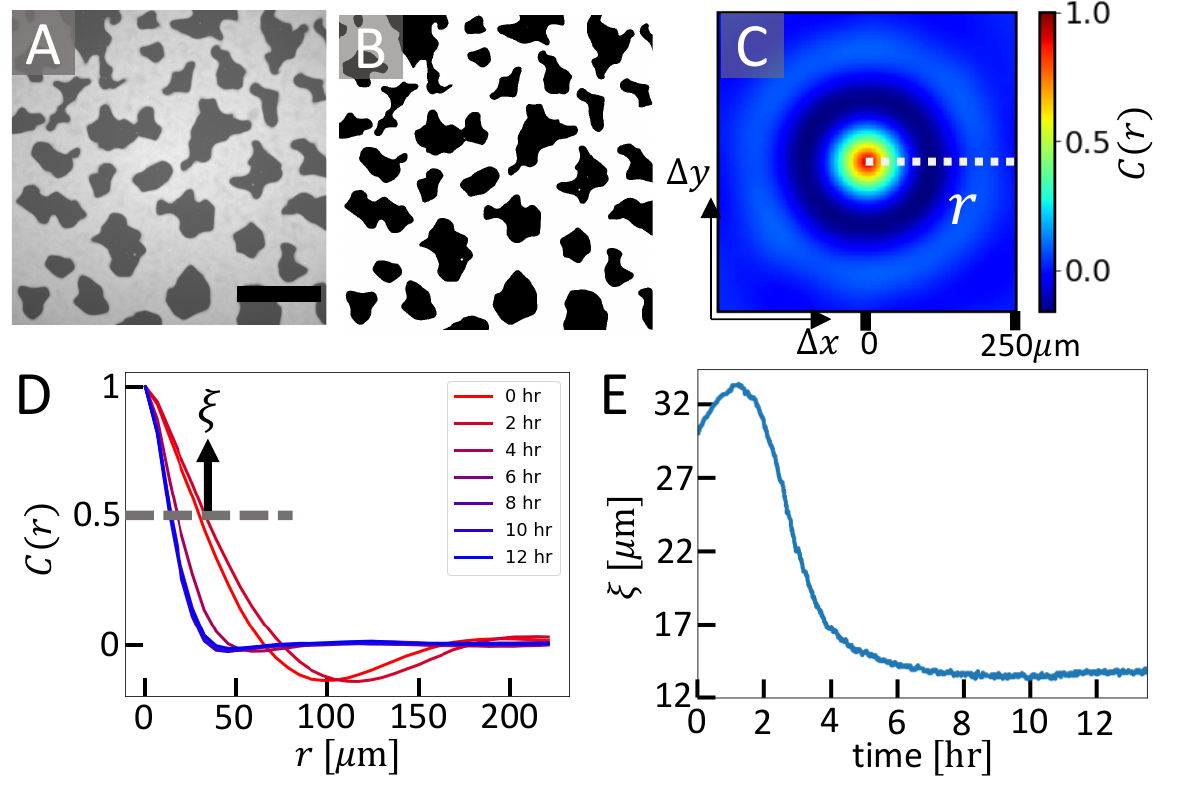}
\caption{\label{fig:correlation} \textbf{Measurement of the correlation length. } (\textbf{A}) Dextran fluorescence image 1.5 hr after the start of the experiment. Only part of the field of view is shown. Scale bar 300 $\mu$m. (\textbf{B}) Pixels in dextran-rich regions are assigned a value of $1$, and those in the PEG-rich regions are assigned a value of $-1$. (\textbf{C}) The autocorrelation matrix of (B). (\textbf{D}) The radial autocorrelation at 6 time points, obtained from (C) by azimuthal averaging around the origin. The correlation length $\xi$ is defined as the distance at which the autocorrelation function equals 0.5.  (\textbf{E}) Evolution of $\xi$ in time. KSA concentration 235nM. }
\end{figure}

\clearpage

\begin{figure}[t]
\centering
\includegraphics[width=\textwidth]{
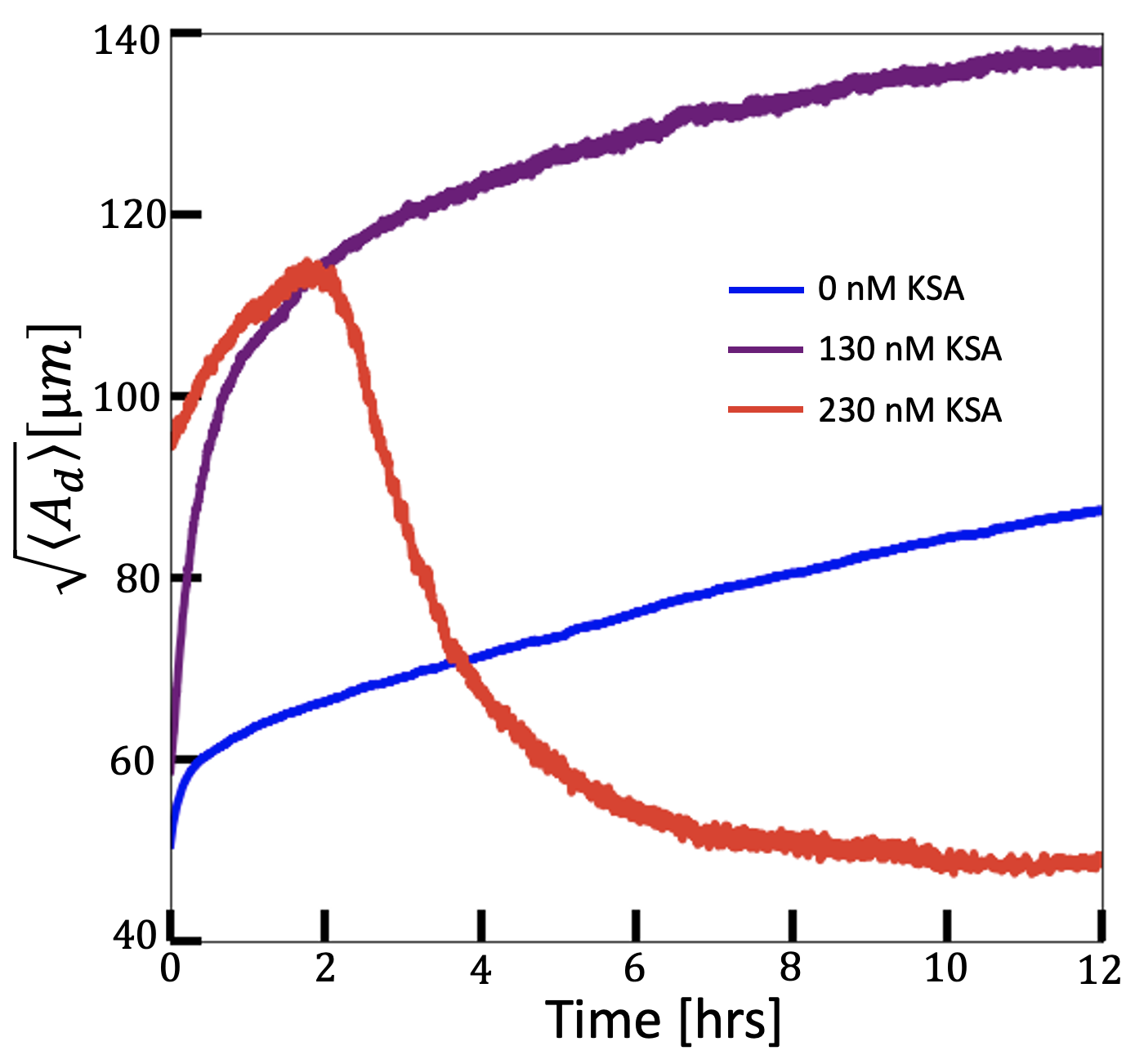}
\caption{\label{fig:MeanDropSize} \textbf{Time evolution of average PEG-rich droplet size.} At each time point, all PEG droplets in a dextran-fluorescence image were identified. Then, their areas were averaged and the average droplet size was defined as the square root of the average area. At 130 nM KSA, droplet coarsening was enhanced. At 230 nM KSA, the mean droplet size peaked around 2 hours, and then entered a dynamic steady state characterized by constant average droplet size.  
}
\end{figure}

\clearpage

\begin{figure}[t]
\centering
\includegraphics[width=\textwidth]{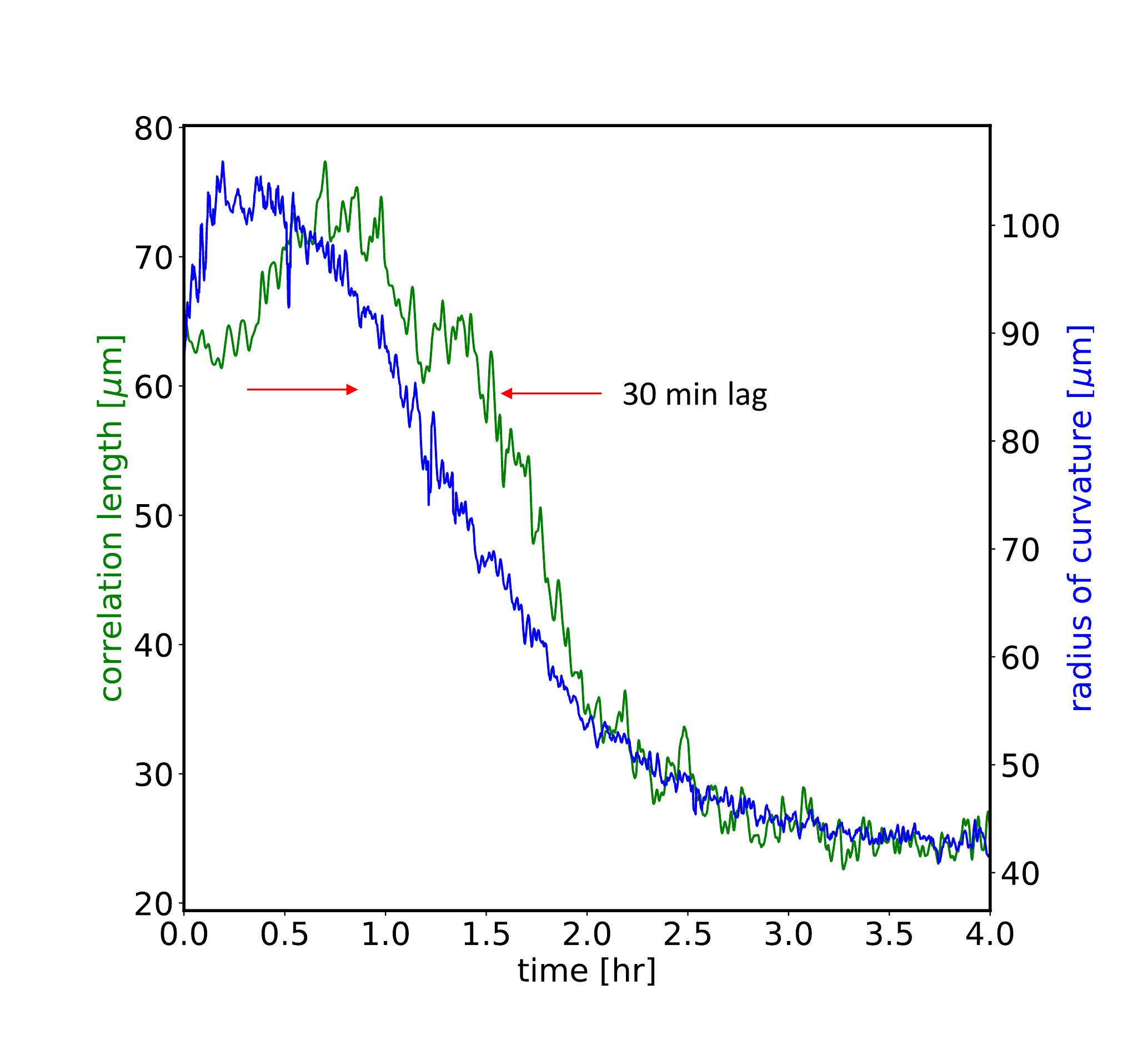}
\caption{\label{fig:correlation_and_curvature} \textbf{Early evolution of correlation length and inverse curvature for 230 nM KSA.} In the first 2 hours of the experiment, correlation length development lagged behind that of the average interface curvature. Subsequently, the rate of change of both quantities coincided as the system approached the steady state.}
\end{figure}

\clearpage

\begin{figure}[t]
\centering
\includegraphics[width=\textwidth]{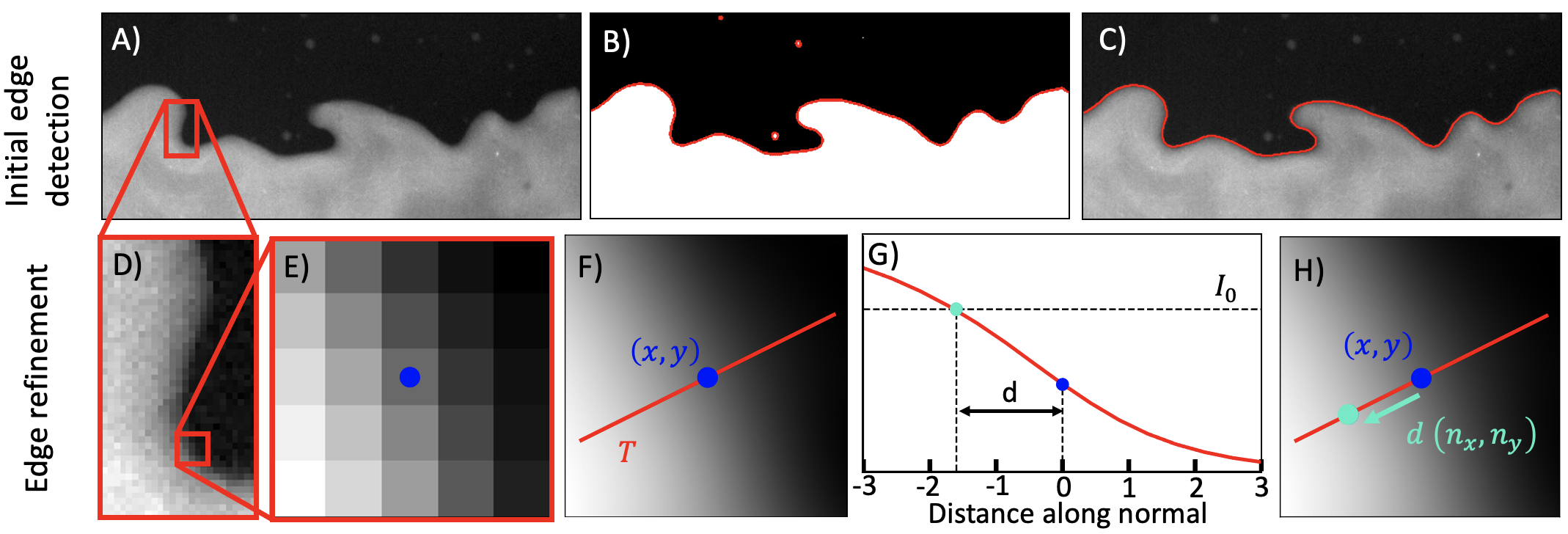}
\caption{\label{fig:InterfaceDetection} \textbf{Detection of bulk-phase-separated interfaces.} (\textbf{A}) Dextran fluorescence image. (\textbf{B}) The thresholded image, with edges detected using the Sobel filter shown in red. (\textbf{C}) After skeletonizing and pruning the largest component, the edge roughly contours the interface. (\textbf{D}) A small section of the interface. (\textbf{E}) A magnified region surrounding a point $(x, y)$ on the interface. (\textbf{F}) Image intensity in (E) is interpolated using a 2D spline (grayscale). Red line is parallel to the local normal $(n_x,n_y)$ to the interface. (\textbf{G}) Interpolated image intensity along the local normal. The distance $\mathrm{d}$ along the normal is defined where the intensity is equal to the threshold. (\textbf{H}) The interface position is defined to be $(x, y) + d (n_x, n_y)$ }
\end{figure}

\clearpage

\begin{figure}[t]
\centering
\includegraphics[width=\textwidth]{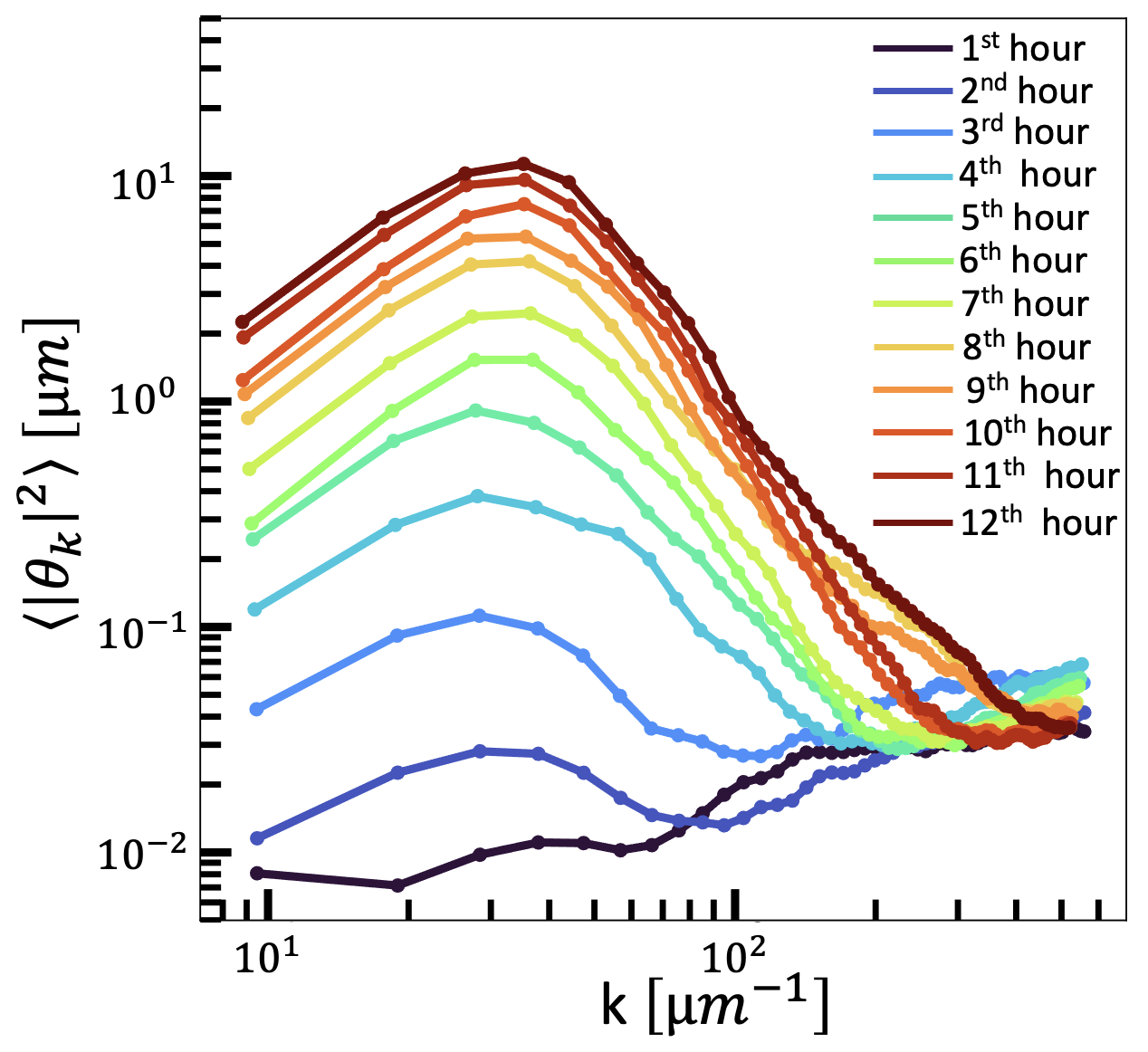}
\caption{\label{fig:TimePSD} \textbf{Tangent angle power spectrum as a function of time, averaged over one-hour intervals after sample preparation.} Although the amplitude of fluctuations increases over time, after $\approx$ 6 hours, the shape of the fluctuation spectrum remains nearly constant. 
}
\end{figure}
\clearpage

\begin{figure}[t]
\centering
\includegraphics[width=\textwidth]{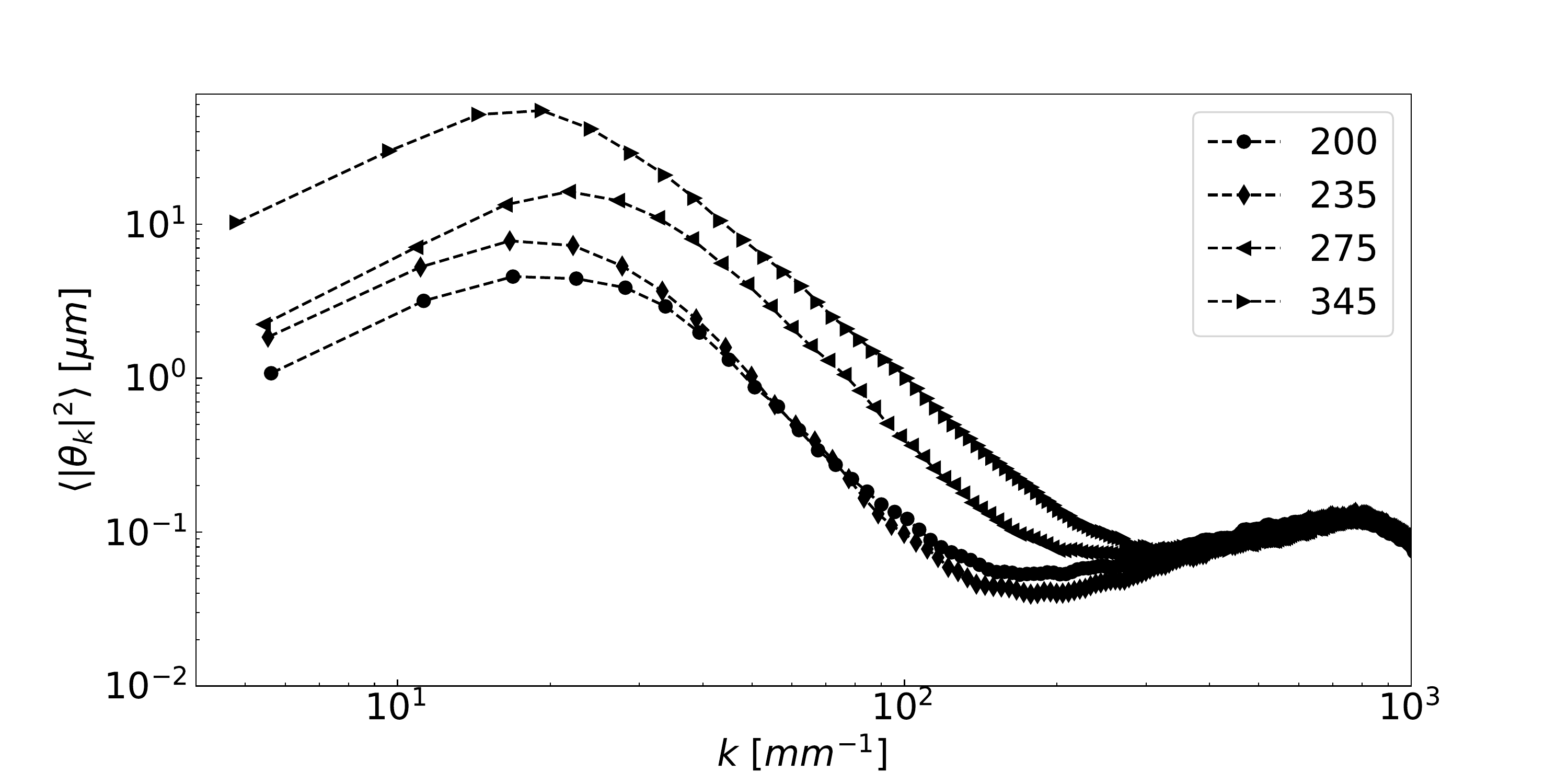}
\caption{\label{fig:NoisePSD} \textbf{Experimental tangent angle spectra extended to $k=10^3$ mm$^{-1}$.} Tangent angle measurement noise increases as $\sim k^2$, resulting in a secondary peak at $k\sim 8\cdot 10^2$ mm$^{-1}$ where all spectra overlap. Legend denotes KSA concentrations in units of nM. 
}
\end{figure}

\clearpage

\begin{figure}[t]
\centering
\includegraphics[width=\textwidth]{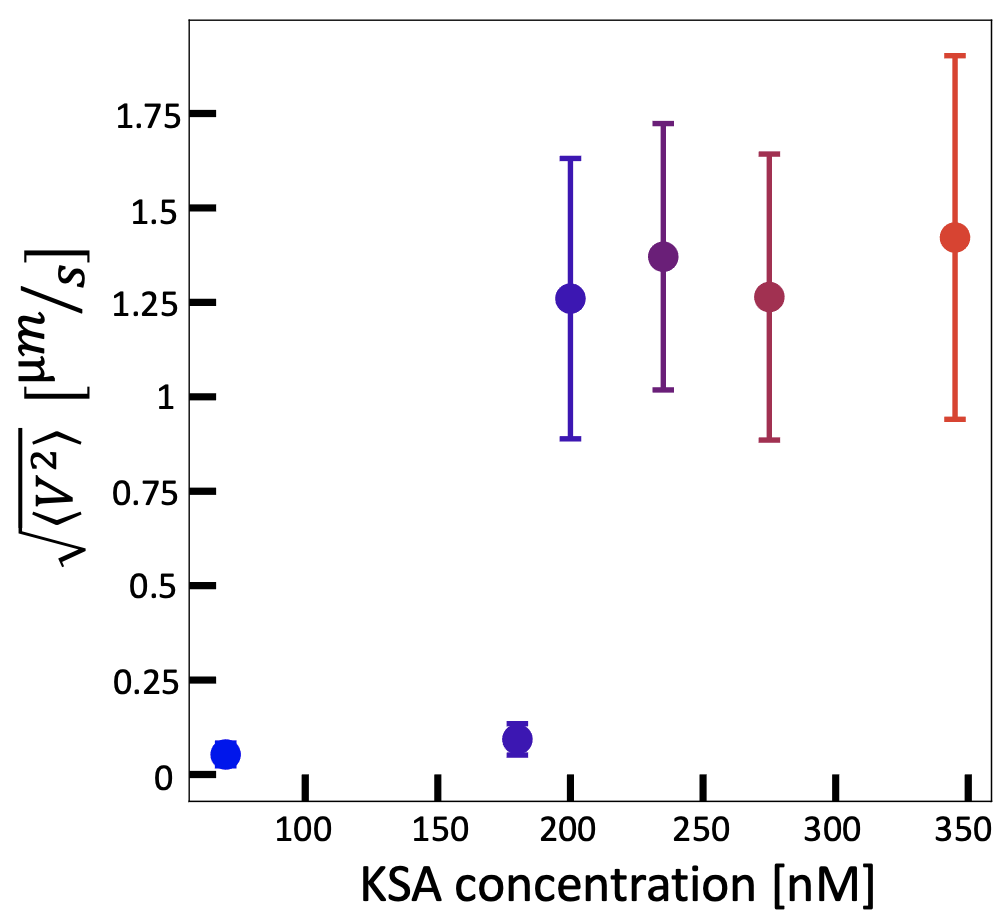}
\caption{\label{fig:VelKSA} \textbf{Root-mean-square velocity of active phase as a function of kinesin concentration averaged from 6 to 8 hours after sample preparation.} Below 200 nM KSA, the speed decreases significantly. Above 200 nM KSA, the speed is approximately constant. Error bars show standard deviation.  
}
\end{figure}

\clearpage

\begin{figure}[t]
\centering
\includegraphics[width=\textwidth]{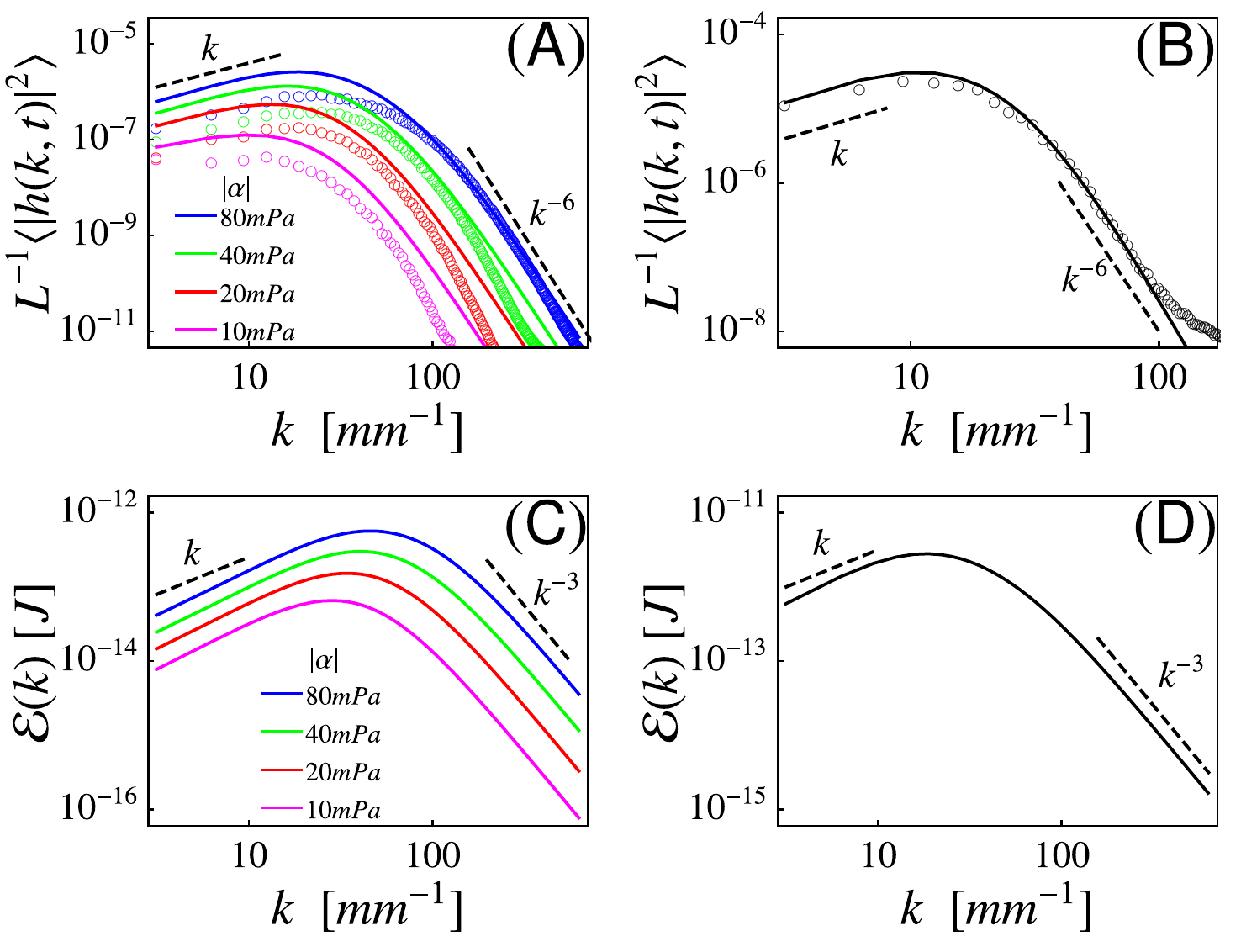}
\caption{\label{fig:pEkHk} \new{\textbf{Fluctuation of active interfaces at different activities.} (\textbf{A}-\textbf{B}) Fluctuation spectra of interfaces from (A) simulations and (B) experiment. Solid lines are theoretical values calculated using Eq. \eqref{eq:PSHa}, while circles are height spectra extracted from either simulations or experiments. For simulations,  $\sigma_{rms}^2=\alpha^2\bar{S}^2/8$, where $\bar{S}$ is the mean nematic order parameter at the steady state. The correlation legnth $\ell_a$, correlation time $\tau_a$, and $\bar{S}$ were measured from simulations. Since all other parameters are known, no fitting parameter is used here. (\textbf{C}-\textbf{D}) Energy spectra $\mathcal{E}(k)$ calculated using \eqref{eq:EkTheory} with parameters from (C) simulations and (D) experiment. The 200 nM KSA data set is used to obtain results in panels B and D. We used the following parameters to calculate the energy and interface fluctuation spectra: interfacial tension $\gamma=0.177\ \mu$N/m, density difference $\Delta \rho=8.9$ kg/m$^3$, viscosity $\eta=25$ mPa$\cdot$S, friction $\gamma_v=100$ MPa. The correlation length and time of active stress $\ell_a=50\ \mu$m and $\tau_a=80$ sec were used in (B) and (D). Magnitude of active stress $\sigma_{rms}$ is used as a fitting parameter, and the best fit gives $\sigma_{rms}=2.47$ mPa.}}
\end{figure}

\clearpage

\begin{figure}[t]
\centering
\includegraphics[width=\textwidth]{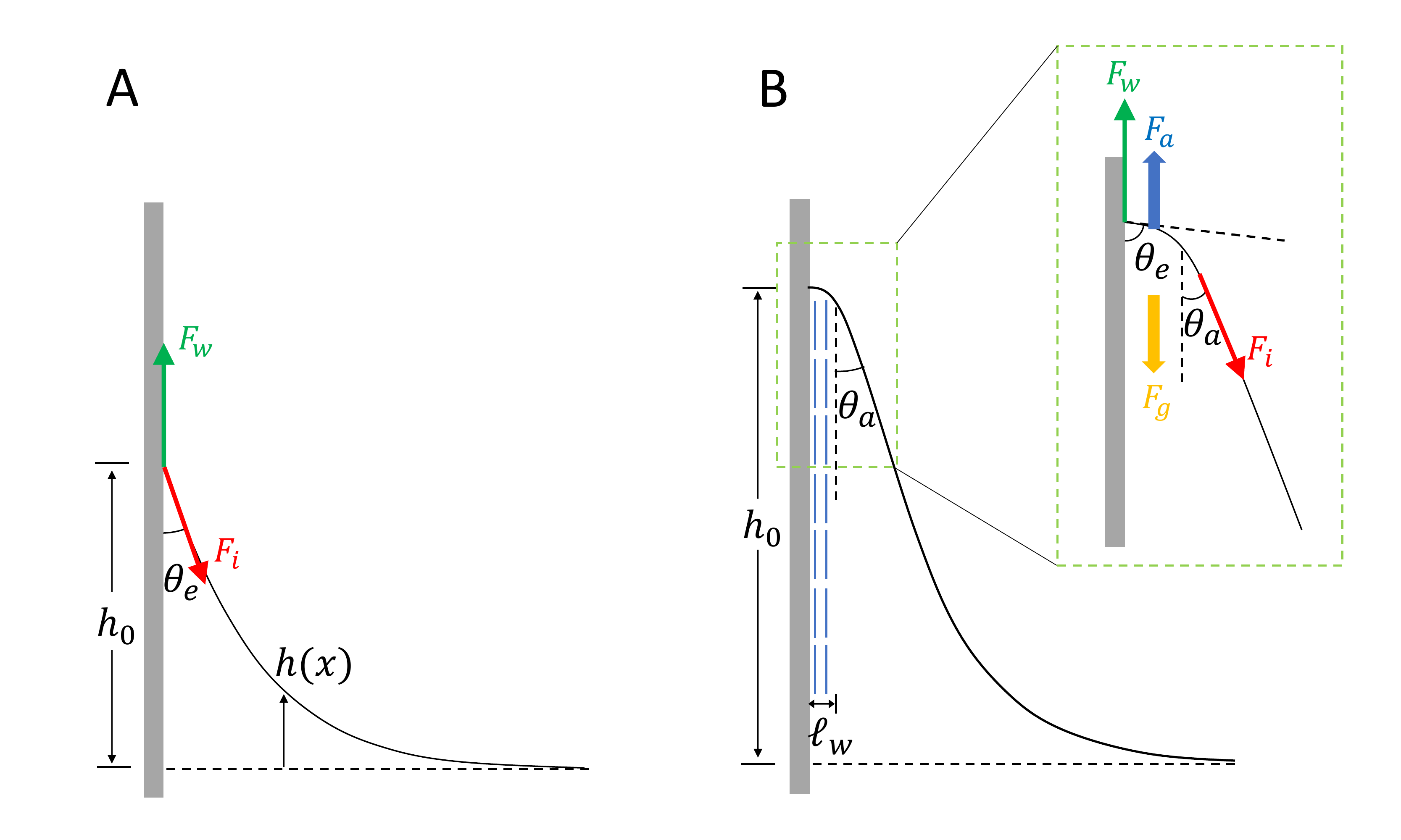}
\caption{\label{fig:wet-sket} \mcm{(\textbf{A}) Sketch of the wetting profile of a passive liquid-air interface defined by $y=h(x)$. The wall tension $\gamma_w$ is determined by $\gamma_w=\gamma_{wall-air}-\gamma_{wall-liquid}$. (\textbf{B}) Sketch of activity-induced wetting. The inset shows the geometry and \red{force balance} close to the contact point.} }
\end{figure}

\clearpage

\begin{figure}[t]
\centering
\includegraphics[width=\textwidth]{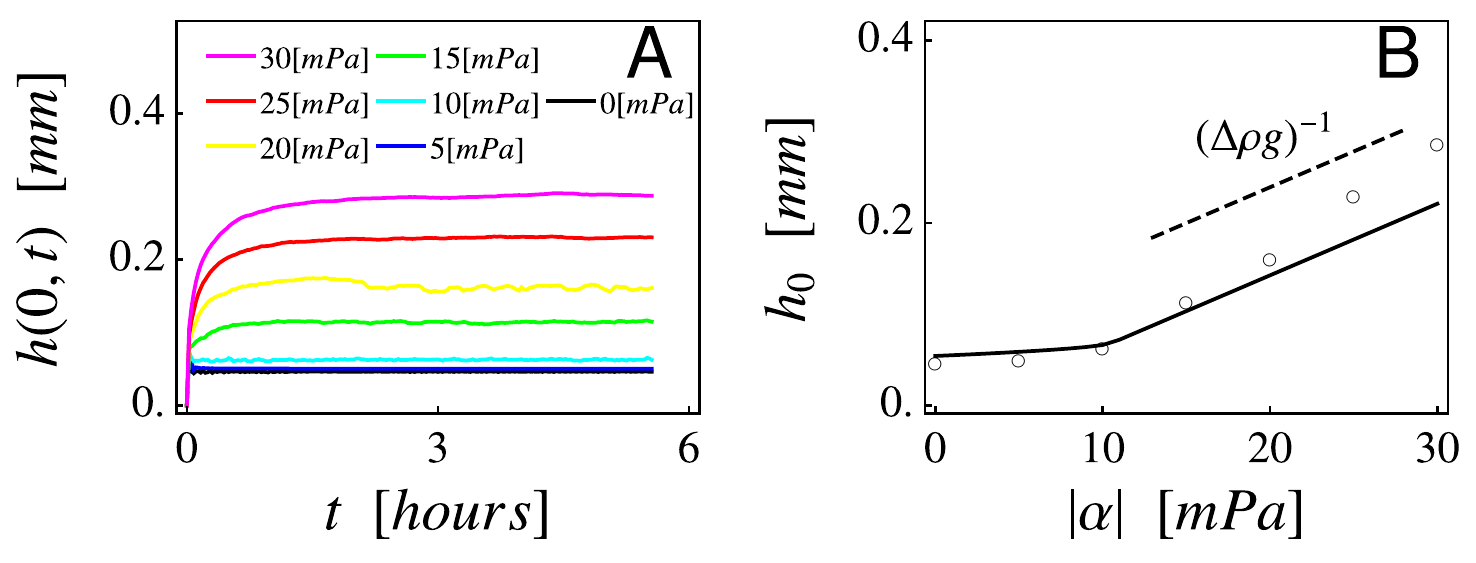}
\caption{\label{fig:h-wet} {\bf Evolution of height of interface contact point with activity.}  (\textbf{A}) Height of contact point as a function of time at different activities \mcm{obtained from numerical simulations of the continuum model.} Since there are two vertical walls, we use the average height of the contact points at the two walls. (\textbf{B}) Steady state height of contact point $h_0$ as a function of activity. The circles are obtained from simulations as $h_0=(t_s-t_0)^{-1}\int_{t_0}^{t_s}h(0,t)dt$, with $t_0=7200 S$ and $t_s=20000 S$. The solid line shows the theoretical value given in  Eq.~\eqref{eq:EqTheta1} with $\ell_w=2.5 \mu m$. The dashed line shows the predicted slope $1/\Delta \rho g$. \mcm{In both figures we have used} a passive contact angle $\theta_e=10$ degree.}
\end{figure}

\clearpage
\begin{figure}[t]
\centering
\includegraphics[width=\textwidth]{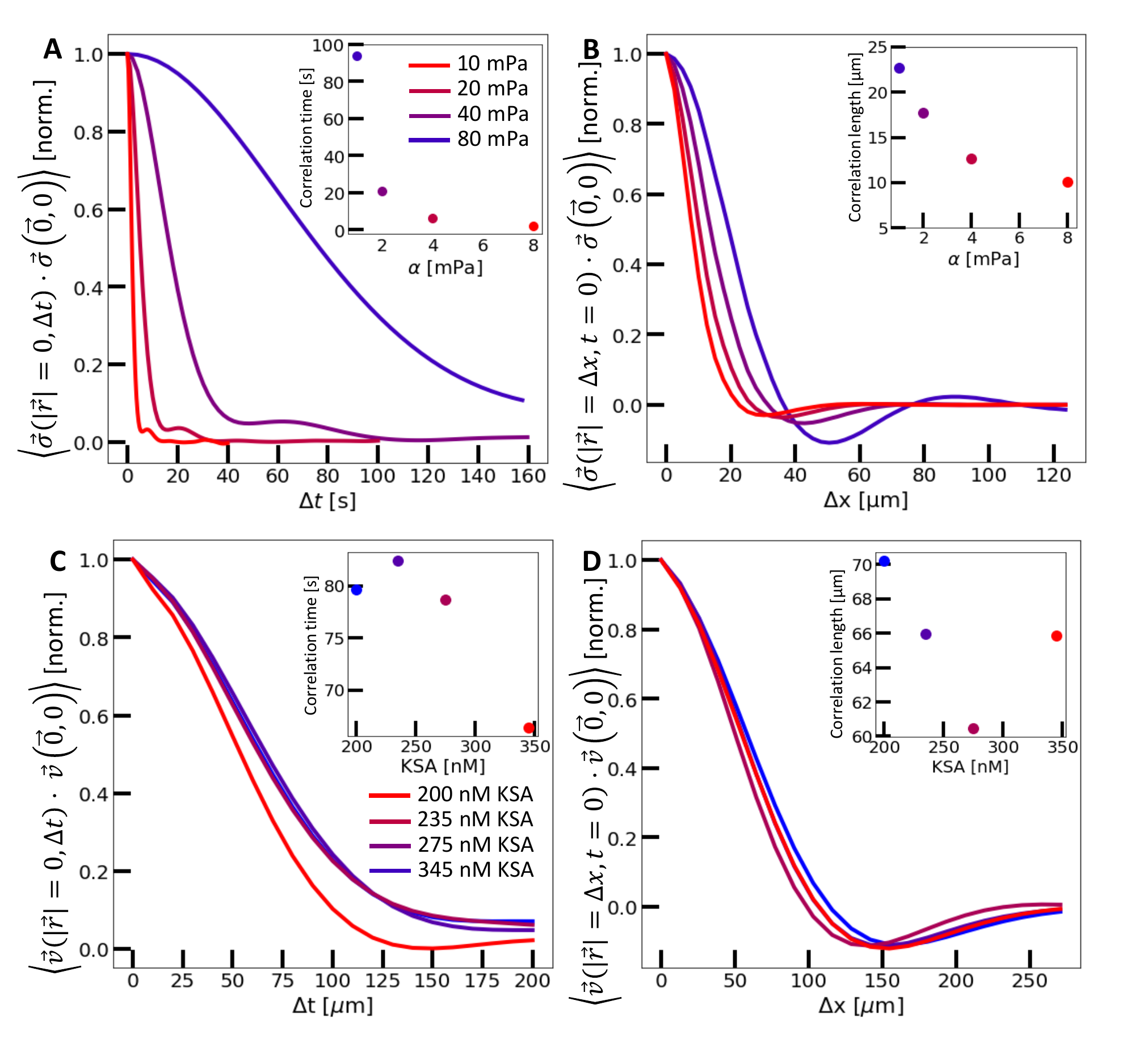}
\caption{\label{fig:bulk-correlations} {\bf Measurement of bulk spatiotemporal autocorrelations in simulations and experiments.} (\textbf{A}) Normalized time correlation of active stresses in numerical simulations. Inset: Correlation time decreases with increasing activity. (\textbf{B}) Normalized space correlation of active stress in numerical simulation. Inset: Correlation length decreases with increasing activity. (\textbf{C}) Normalized time correlation of velocity in experiment. Inset: Correlation time shows little variation with activity. (\textbf{D}) Normalized space correlation of velocity in experiment. Inset: Correlation length shows little variation with activity. Velocity (stress) autocorrelations were computed by averaging the scalar products of bulk velocity (stress) fields with their displacement in time $\Delta t $ and space $\Delta x$. Correlation times and length were defined where the correlation curve reaches $1/e$.}
\end{figure}

\clearpage
%\clearpage
%\begin{figure}[h]
%\centering
%\includegraphics[width=0.6\textwidth]{coalescence.pdf}
%\caption{\label{fig:coalescence} \textbf{Interfacial tension determination from droplet coalescence.} (A) Time sequence of two coalescing dextran-rich droplets. Dextran phase was prepared by mixing 4.2\% w/w  dextran with 0.9\% w/w PEG. PEG phase had 2.5 \% w/w PEG and no dextran. Alexa-488 fluorescently labelled 2MDa dextran was added to the dextran phase. The two phases were mixed at the ratio 1:9 (dextran phase:PEG phase), and flowed into a chamber containing fluorinated oil with surfactant (HFE 7500, 3M). Droplets sedimented on the oil-water interface, where they were observed via confocal microscopy. W and L define the width and length of the coalescing droplet respectively. (B) Decay of the geometric ratio A over time. Black line is an exponential fit $A=e^{-t/\tau}$ used to extract the relaxation time $\tau$. (C) Relaxation times from 7 coalescence events versus equilibrium droplet radius. Black line is a linear fit.}
%\end{figure}
%\clearpage

\textbf{Supplementary movie captions}
\newline

\textbf{Movie S1:} Phase separation dynamics at 0 nM KSA concentration. Images taken at 4 min interval (top panel). Grayscales depict dextran-488 fluorescence, highlighting the active phase. Chamber thickness, 30~$\mu$m. Evolution of the correlation length (bottom panel). 

\textbf{Movie S2:} Phase separation dynamics at 135 nM KSA concentration. Images taken at 4 min interval (top panel). Grayscales depicts dextran-488 fluorescence, highlighting the active phase. At this kinesin concentration, the passive PEG-rich regions are advected by the active phase, enhancing the coarsening dynamics. Chamber thickness, 30~$\mu$m. Evolution of the correlation length (bottom panel). 

\textbf{Movie S3:} Phase separation dynamics at 230 nM KSA concentration. Images taken at 2 min intervals. Grayscales depicts dextran-488 fluorescence, highlighting the active phase. Droplets undergo fusion and fission events that suppress coarsening. Chamber thickness, 30~$\mu$m. Evolution of the correlation length (bottom panel). 

\textbf{Movie S4:} High resolution images of phase separation dynamics at 230 nM KSA concentration, taken at 45 sec intervals. Grayscales depicts dextran-488 fluorescence, highlighting the active phase. Splitting and merging of droplets are clearly visible. Chamber thickness, 30~$\mu$m. Evolution of the correlation length (bottom panel).

\textbf{Movie S5:} Fluctuations of an interface at 345 nM KSA concentration taken at a 15 sec interval with a macro lens (Canon M50, EFS 60mm f/2.8). Fluctuations are several tens of microns in height, and are clearly visible to the naked eye due to the refractive index mismatch between the phases.

\textbf{Movie S6:} Time series of active interfaces taken at 10 sec intervals, and at three KSA concentration. Detected interfaces (red lines). Interfacial fluctuation amplitude grows with KSA concentration. Chamber thickness, 60~$\mu$m. Travelling disturbances of the interface are indicated with arrows. Grayscales represent dextran-488 fluorescence. Scale bar, 350 $\mu$m.

\textbf{Movie S7:} Numerical steady-state interfacial fluctuations. Legends denote activities $|\alpha|$. Grayscales depict the nematic order parameter, $S$. Scale bar, 100~$\mu$m. 2 second time step.

\textbf{Movie S8: } Dynamics of the wetting profiles of active interfaces in contact with acrylamide-coated glass walls. Images were taken at 10 sec intervals. Grayscales depict dextran-488 fluorescence.

\textbf{Movie S9:} Numerical active wetting profiles. Legends denote activities $|\alpha|$. Equilibrium contact angle, $\theta_e = 10^\circ$. Grayscales depict the nematic order parameter $S$. Scale bar, 50~$\mu$m. 20 second time step.    

\clearpage

\end{document}